\documentclass[fleqn]{aa}
\usepackage{amsmath,amssymb,graphics,times}

\allowdisplaybreaks

\newfont{\eufont}{eufm10}
\def\eu #1{\mbox{\eufont #1}}

\newcommand{\Teff}{\mbox{$T_\mathrm{eff}$}}

\newcommand{\vr}{{v_r}}
\newcommand{\pr}[1]{{#1}_{\mathrm{p}}}
\newcommand{\io}[1]{{#1}_{\mathrm{i}}}
\newcommand{\el}[1]{{#1}_{\mathrm{e}}}
\newcommand{\erf}{\mathrm{erf}}
\newcommand{\zav}[1]{\left(#1\right)}
\newcommand{\hzav}[1]{\left[#1\right]}
\newcommand{\mzav}[1]{\left\{#1\right\}}
\newcommand{\xa}{\io{\eu Y}}
\newcommand{\vth}{v_{\mathrm{th}}}
\newcommand{\vthi}{v_{\mathrm{th,i}}}
\newcommand{\rhvez}{R_{*}}
\newcommand{\txi}{\boldsymbol{\xi}}
\newcommand{\krat}{\!\cdot\!}
\newcommand{\pd}[2]{\frac{\partial #1}{\partial #2}}
\newcommand{\pul}{\frac{1}{2}}
\newcommand{\DeltanuD}{\Delta\nu_\mathrm{D}}
\newcommand{\GO}{Gayley-Owocki}
\newcommand\zm[1]{#1}

\newcommand{\obrazek}[3]{%
    \begin{figure}
    \resizebox{\hsize}{!}{\includegraphics{#1}}
    \caption[]{#2}
    \label{#3}
    \end{figure}
}

\newcommand{\obrazekd}[4]{%
    \begin{figure}
    \resizebox{\hsize}{!}{\includegraphics{#1}}
    \resizebox{\hsize}{!}{\includegraphics{#2}}
    \caption[]{#3}
    \label{#4}
    \end{figure}
}

\begin{document}

\title{Multicomponent radiatively driven stellar winds}
\subtitle{II. \GO\ heating in multitemperature winds of OB stars}

\titlerunning{Multicomponent radiatively driven stellar winds II.}

\author{Ji\v{r}\'{\i} Krti\v{c}ka\inst{1,2} \and
        Ji\v{r}\'{\i} Kub\'at\inst{2}}
\authorrunning{J. Krti\v{c}ka, J. Kub\'at}

\offprints{J. Krti\v{c}ka,\\ \email{krticka@physics.muni.cz}}

\institute{\'Ustav teoretick\'e fyziky a astrofyziky P\v{r}F MU,
            Kotl\'a\v{r}sk\'a 2, CZ-611 37 Brno, Czech Republic 
           \and
           Astronomick\'y \'ustav, Akademie v\v{e}d \v{C}esk\'e
           republiky, CZ-251 65 Ond\v{r}ejov, Czech Republic}

\date{Received 29 May 2001}

\abstract{We show that the so-called \GO\ (Doppler) heating is
important for the temperature structure of the wind of main sequence
stars cooler than the spectral type O6.
The formula for \GO\ heating is derived directly from the Boltzmann
equation as a direct consequence of the dependence of the driving force
on the velocity gradient.
Since \GO\ heating deposits heat directly to the absorbing ions, we also
investigated the possibility that individual components of the
radiatively driven stellar wind have different temperatures.
This effect is negligible in the wind of O stars, whereas a significant
temperature difference takes place in the winds of main sequence B stars
for stars cooler than B2.
Typical temperature difference between absorbing ions and other flow
components for such stars is of the order $10^3\,\mathrm{K}$.
However, in the case when passive component falls back onto the star the
absorbing component reaches temperatures of order $10^6\,\mathrm{K}$,
which allows for emission of X-rays.
\\
Moreover, we compare our computed terminal velocities with the observed
ones.
We found quite good agreement between predicted and observed terminal
velocities.
The systematic difference coming from the using of the so called
``cooking formula'' has been removed.
   \keywords{Hydrodynamics --
             stars:  mass-loss  --    
             stars:  early-type  --   
             stars:  winds}
}

\maketitle

\section{Introduction}

Since the foundation of the theory of radiatively driven stellar winds
by Lucy \& Solomon (\cite{LS70}) and  Castor, Abbott \& Klein
(\cite{cak}, hereafter CAK)  many of the initial assumptions introduced
by this authors were examined.
To the most important ones belong
the radial streaming approximation (Friend \& Abbott \cite{fa},
Pauldrach et al. \cite{ppk}), the wind stability (Abbott \cite{abb},
Owocki \& Rybicki \cite{ornest}), the limitations of the Sobolev
approach (Poe et al. \cite{poc},
Owocki \& Puls \cite{opsob}), the thermal structure of the wind (Drew
\cite{drewmoc}) and many others.

Another important assumption, studied already at the dawn of the
radiatively driven stellar wind theory by Castor, Abbott \& Klein
(\cite{cak76}) is the condition of the one-component flow.
They discussed encounters which transfer momentum received by absorbing
ions (typically C, N, O, etc.) to passive, nonabsorbing ions, mainly
hydrogen and helium.
They showed that for the high-density winds such encounters are not
important for the overall dynamics of the wind and that high-density
winds can be considered as one-component.
However, for the low-density winds Springmann \& Pauldrach
(\cite{treni}, hereafter SP) showed that momentum transfer between
absorbing and nonabsorbing plasma influences the wind thermal balance
and even the wind dynamics.
Thus, for the low-density winds the flow is essentially multicomponent.
They proposed that the so-called "ion-runaway" may occur.
Based on the simplified theory of the multicomponent flow many of
interesting results occurred.
Porter \& Drew (\cite{iontdisk}) re-examined model of wind-compressed
disk in the presence of dynamical decoupling of absorbing ions and
passive plasma, Porter \& Skouza (\cite{obalka}) showed the possibility
of formation of pulsating shells around stars with low-density
radiatively driven wind, and Hunger \& Groote (\cite{HuGr}) explained
the H/He abundance anomalies in Bp stars on the basis of helium
decoupling.

First detailed numerical models of multicomponent radiatively driven
stellar winds were presented by Babel (\cite{babela}, \cite{babelb}).
However, Krti\v{c}ka \& Kub\'at (\cite{kk}, hereafter KK0) showed that
due to the functional dependence of the radiative force decoupling does
not occur.
Moreover, Krti\v{c}ka \& Kub\'at (\cite{mydvai}, hereafter KKI) using
nonisothermal multicomponent models concluded that winds of B stars are
frictionally heated in such amount that the possibility of decoupling of
absorbing ions from the passive plasma is excluded.

The solar wind is well-known to posses large temperature differences
between electrons and protons.
Such differences were obtained also by B\"urgi (\cite{burko}), who used
the three-component models of the solar wind.
So the natural question arises, whether similar temperature differences
exist in the radiatively driven stellar wind or, in other words, whether
the assumption of equal temperatures of all wind components is
acceptable.
In this paper we intend to answer this question.

Any effect which deposits heat separately to individual component of
the flow may influence our results.
Thus, we shall include the effect of Doppler heating, introduced in the
stellar wind domain by Gayley \& Owocki (\cite{dop}, hereafter GO). 
Because it arises from the dependence of the radiative force on the
velocity via the Doppler effect it deposits heat directly to the
absorbing ion component and thus, it can trigger the temperature
difference between absorbing and passive ions.

Proper treatment of ionization balance may be important for correct
description of decoupling of individual components of the flow.
Thus, we decided to compute electrical charges of individual components
using adequate ionization balance formulas.

%%%%%%%%%%%%%%%%%%%%%%%%%%%%%%%%%%%%%%%%%%%%%%%%%%%%%%%%%%%%%%%%%%%%%%%%
%                        Doppler heating
%%%%%%%%%%%%%%%%%%%%%%%%%%%%%%%%%%%%%%%%%%%%%%%%%%%%%%%%%%%%%%%%%%%%%%%%

\section{Boltzmann equation with velocity-dependent force}

The procedure of the derivation of the hydrodynamic equations from the
Boltzmann equation for particle distribution function $F_s$ of the
particle $s$ is thoroughly described in a number of textbooks.
However, it is commonly assumed that the Boltzmann equation can be
written in the form (we use the Einstein summation law)
\begin{equation}
\frac{\partial F_s}{\partial t} +
\xi_{sh}\frac{\partial F_s}{\partial x_h} +
\frac{f_{sh}}{m_s}
\frac{\partial F_s}{\partial \xi_{sh}} =
\zav{\frac{dF_s}{dt}}_{\mathrm{coll}},
\end{equation}
i.e. it is assumed that the force $\boldsymbol{f}_s$ is
{\em independent} of velocity, which is acceptable when we consider
gravitational or electrical forces.
However, the latter assumption is not valid for the case of a wind
driven by radiative force coming from line absorption, which is strongly
dependent on a velocity gradient.
Therefore, we rederive here hydrodynamic equations without an
assumption of the force independent of velocity.
In this case, the Boltzmann equation for the one-particle
distribution function $F_s$ of particles of type $s$ is written as 
\begin{equation}
\label{bollep}
\frac{\partial F_s}{\partial t} +\xi_{sh}
\frac{\partial F_s}{\partial x_h} +
\frac{\partial }{\partial \xi_{sh}}\zav{\frac{f_{sh}}{m_s}F_s} =
\zav{\frac{dF_s}{dt}}_{\mathrm{coll}}.
\end{equation}
Here $\xi_{sh}$ 
($h=1,2,3$) are the velocity components of
individual particles of a type $s$ with mass $m_s$, and $f_{sh}$ are the
components of an external force acting on them.
Right-hand side term expresses the effect of collisions.
By a definition, the integral of the distribution function over the
velocity space is a number density $n_s$ of $s$-particles,
\begin{equation}
\label{defn}
n_s=\int d\txi_s F_s.
\end{equation}
Now, the usual way to obtain hydrodynamic  equations is the following.
One multiplies the Boltzmann equation (\ref{bollep}) by multipliers
$m_s$, $m_s\xi_{sh}$ and $m_s\xi_{sh}\xi_{sk}$
and integrates it over the velocity space.
For the discussion of the role of velocity dependent force in the
Boltzmann equation we confine us to the left-hand side of the Boltzmann
equation.
The right-hand side (i.e. the collisional term) remains unaffected by
a presence of such forces,
so we assume in this section that the gas is collisionless, i.e. that
the right hand side of the Boltzmann equation is zero.

\subsection{Continuity equation}

Multiplying the Boltzmann equation (\ref{bollep}) by $m_s$ and
integrating over the velocity space we obtain the continuity equation
\begin{equation}
\label{rovmomkont}
\pd{}{t}\zav{n_s m_s}+\pd{}{x_k}\zav{n_s m_s v_{sk}}=0,
\end{equation}
where $v_{sk}$ are the components of the mean velocity of particles $s$,
and
\begin{equation}
n_s m_s v_{sk}= m_s \int d\txi_s \xi_{sk} F_s.
\end{equation}
Here, the force term disappeared through integration by parts and the
collisional term is also zero.
Apparently, velocity-dependent external force does not change the
continuity equation.

\subsection{Momentum equation}

Multiplying the Boltzmann equation (\ref{bollep}) by $m_s\xi_{sh}$ and
integrating over the velocity space we obtain momentum equation
\begin{multline}
\label{rovmompohc}
\pd{}{t}\zav{n_s m_s v_{sh}}+
\pd{}{x_k}\zav{n_s m_s v_{sh} v_{sk}+p_{s,hk}}-\\*
\int d\txi_s  f_{sh} F_s = 0,
\end{multline}
where $p_{s,hk}$ are the components of the momentum transfer tensor, and
\begin{equation}
n_s m_s v_{sh} v_{sk}+p_{s,hk} = m_s \int d\txi_s \xi_{sh} \xi_{sk} F_s.
\end{equation}
Again, the velocity term was integrated by parts. Note, that due to the
velocity dependence of the external force the term containing $f_{sh}$
cannot be moved in front of the integral.

\subsection{Energy equation}

Multiplying Boltzmann equation (\ref{bollep}) by $m_s\xi_{sk}\xi_{sk}$
and integrating over the velocity space we obtain energy equation 
\begin{multline}
\label{rovmomenergc}
\pd{}{t}\zav{\pul n_s m_s  v_s^2+\frac{3}{2} p_s}+\\*+
\pd{}{x_i}\zav{\pul n_s m_s v_s^2 v_{si}+\frac{3}{2} p_s v_{si}+
v_{sk}p_{s,ki}+\frac{1}{2}p_{s,kki}}-\\*-
\int d\txi_s \xi_{sk}f_{sk}F_s=0,
\end{multline}
where $p_s=1/3 p_{s,kk}$ is the scalar hydrostatic pressure and
the last term was simplified using integration by parts.
The equation for the temperature can be derived by subtracting the
momentum equation (\ref{rovmompohc}) dot-multiplied by $v_{sh}$ from
the Eq. (\ref{rovmomenergc}),
\begin{multline}
\label{rovtepbur}
\pd{}{t}\zav{\frac{3}{2} p_s}+\pd{}{x_i}\zav{\frac{3}{2} p_s v_{si}}+
p_{s,ij}\pd{v_{sj}}{x_i}+\pul\pd{}{x_i}p_{s,jji}+\\*+
v_{sh}\int d\txi_s  f_{sh} F_s-\int d\txi_s \xi_{sh}f_{sh}F_s =0.
\end{multline}
Note that the last two terms cancel if the force does not depend on the
velocity.
For this ``standard'' case the external force does not give any
contribution to net heating.

\subsubsection{\GO\ (Doppler) heating}

Let us explore now the last term on the left hand side of Eq.
(\ref{rovtepbur}) for the case of the force caused by absorption of
radiation in spectral lines.
This force depends on particle velocity through the velocity dependence
of the line absorption coefficient owing to the Doppler effect.
Therefore, Gayley \& Owocki (\cite{dop}, hereafter GO) termed the
heating effect by Doppler heating, but terming it
\GO\ heating (or GO heating in the abbreviated form)
might be more appropriate.
Let us denote the heating term in the {\em comoving} 
fluid-frame as
\begin{equation}
\io Q^{\mathrm{GO}} =
\int d\io\txi \xi_{\mathrm{i},h}f_{\mathrm{i},h}\io F.
\end{equation}
Here the index 
$\mathrm{i}$
stands for ``absorbing ions''.
Although this term is written in the comoving fluid-frame, the same
expression holds in the non-relativistic case also in the
observer frame.
Let us assume complete redistribution and an angle independent
opacity and emissivity in the atomic frame.
However, the emissivity in the comoving fluid-frame is generally angle
dependent due to the Doppler effect.
Therefore, the radiative force acting on an atom in the
comoving fluid-frame is
\begin{equation}\label{zarsila}
f_{\mathrm{i},h} = \frac{m_\mathrm{i}}{c} \int_0^\infty d \nu
\oint d \omega \hzav{\kappa(\vec{n},\nu) I(\vec{n},\nu) - \epsilon(\vec{n},\nu)}
n_h,
\end{equation}
where $\kappa(\vec{n},\nu)$ and $\epsilon(\vec{n},\nu)$ are the
absorption and emission coefficients per unit mass, which can be
expressed as
\begin{subequations}
\begin{align}
\kappa(\vec{n},\nu) &=
\kappa\, \varphi\!\zav{\nu-\frac{\nu_0}{c}\io\txi\krat\boldsymbol{n}},\\
\epsilon(\vec{n},\nu) &=
\epsilon\, \varphi\!\zav{\nu-\frac{\nu_0}{c}\io\txi\krat\boldsymbol{n}},
\end{align}
\end{subequations}
where $\varphi(\nu)$ is the absorption (emission) profile in the atomic
frame.
After some rearrangement we obtain
\begin{multline}
\io Q^{\mathrm{GO}} =
\frac{\io m}{c} \int_{0}^{\infty}d\nu \oint d\omega\, 
\hzav{\kappa I(\boldsymbol{n},\nu) - \epsilon} \times\\*\times
\int d\io\txi\,  \io\txi\krat\boldsymbol{n} \io F
\varphi\zav{\nu-\frac{\nu_0}{c}\io\txi\krat\boldsymbol{n}}.
\end{multline}
If for the calculation of the last integral we choose one of the 
velocity axes parallel to the direction of $\boldsymbol{n}$, then the 
integrals over other two axes vanishes (we integrate odd function) and
the \GO\ heating formula becomes
\begin{multline}
\io Q^{\mathrm{GO}} =
\frac{\io m}{c} \int_{0}^{\infty}d\nu \oint d\omega\, 
\hzav{\kappa I(\boldsymbol{n},\nu)-\epsilon}
\times \\ \times
\int d\io\xi  \,\io\xi\io F(\io\xi)
\varphi
\zav{\nu-\frac{\nu_0}{c}\io\xi}.
\end{multline}
where 
$\io\xi$
is a velocity component in the direction of
$\boldsymbol{n}$.
If we rewrite photon-line-of-sight velocity component as 
$\io\xi=w\vthi$
and use frequency displacement from the line center in Doppler units
$x=\zav{\nu-\nu_0}/\DeltanuD$, then the GO heating takes the form
\begin{multline}
\io Q^{\mathrm{GO}} =
\frac{\io m}{c} \int_{-\infty}^\infty dx
\oint d\omega
\hzav{\kappa \tilde{I}(\boldsymbol{n},x)-\Delta\nu_D\epsilon}
\times\\* \times
\int_{-\infty}^\infty \vthi^2\,dw\,w\,
\io F(\vthi w)\,\psi(w-x),
\end{multline}
where we introduced the intensity $\tilde{I}(\boldsymbol{n},x)$ as
$\tilde{I}(\boldsymbol{n},x)\, dx = I(\boldsymbol{n},\nu)\, d\nu$ and
the function
\begin{equation}
\psi(x)=\varphi(x\DeltanuD)
\end{equation}
is normalized according to Castor (\cite{cassob}) as
\begin{equation}
\int_{-\infty}^{\infty} \psi(x) dx=1.
\end{equation}
Note that the thermal speed $\vthi$ is really ionic because it comes
from the velocity distribution of absorbing ions (shall not be
interchanged with $\vth$, which comes from normalization of force
multipliers).
We neglect absorption in the resonance wings of the profile and
approximated $\psi(w-x)\approx\delta(w-x)$.
Finally, 
we assume that the velocity distribution is
given by the Maxwellian velocity distribution,
\begin{displaymath}
\io F(\vthi w) =\frac{\io n}{\vthi \sqrt\pi}\,e^{-w^2}.
\end{displaymath}
\zm{The latter assumption was made purely due to simplicity reasons.
Relaxing it could lead to interesting effects especially if
the number of collisions is not sufficient to maintain an equilibrium
(cf., e.g., Scudder \cite{scudder}, Cranmer \cite{cranmer}).
However, we postpone the analysis of the non-Maxwellian effects to a
future paper.}

Thus, the GO heating formula takes the form of
\begin{multline}
\io Q^{\mathrm{GO}} =
\frac{\io m \io n \vthi}{c}
\oint d\omega\, \int_{-\infty}^{\infty} dx
\,x \phi(x)
\times\\* \times
\hzav{\kappa \tilde I(\boldsymbol{n},x)-\DeltanuD\epsilon},
\end{multline}
where
\begin{equation}\label{absprof}
\phi(x)=\frac{1}{\sqrt\pi}e^{-x^2}.
\end{equation}
Due to the symmetry of the absorption profile (\ref{absprof})
the product $x \phi(x)$ is an odd function and, thus, after integration
over $x$ vanishes.
This means that in the case of complete redistribution the process of
emission gives no {\em direct} contribution to
GO
heating, i.e.
\begin{multline}
\io Q^{\mathrm{GO}} =
\frac{\io m \io n \vthi}{c}
\oint d\omega\, \int_{-\infty}^{\infty} dx
\,x \phi(x)
\kappa \tilde I(\boldsymbol{n},x).
\end{multline}
In the static medium the product 
$x \phi(x) \tilde I(\boldsymbol{n},x)$ is an odd function of $x$,
thus, there is no
GO
heating effect in static stellar atmospheres.
In the particular case of a spherically-symmetric stellar wind the
expression for the \GO\ heating takes the form of
\begin{equation}
\label{qdopob}
\io Q^{\mathrm{GO}} = \frac{2\pi\kappa\io\rho\vthi}{c}
\int_{-1}^1d\mu \int_{-\infty}^{\infty} dx\, x \phi(x) \tilde I(\mu,x),
\end{equation}
where $\mu=\cos\theta$, which was actually used by GO.

\subsubsection{Formula for
\GO\
heating in the stellar wind domain}

In the case of a two-level atom without continuum the solution of the
transfer equation in the Sobolev approximation is (Rybicki \& Hummer
\cite{rybashumrem}, Owocki \& Rybicki \cite{OR2}, GO)
\begin{equation}
\label{isob}
\tilde I(\mu,x)=\tilde I_c
\mzav{\frac{\beta_c}{\beta}+\hzav{D(\mu)-
\frac{\beta_c}{\beta}}
e^{-\tau_\mu\Phi(x)}},
\end{equation}
where $\tilde I_c$ is the core intensity,
$D(\mu)$ is unity for $\mu>\mu_*$ and
zero otherwise ($\mu_*=\zav{1-R_*^2/r^2}^{1/2}$), core penetration and
escape probabilities are given by
\begin{align}
\label{betac}
\beta_c&=\frac{1}{2}\int_{\mu_*}^{1}d\mu\frac{1-e^{-\tau_\mu}}{\tau_\mu},\\
\label{beta}
\beta&=\frac{1}{2}\int_{-1}^{1}d\mu\frac{1-e^{-\tau_\mu}}{\tau_\mu},
\end{align}
respectively, and
\begin{equation}
\Phi(x)=\int_{x}^{\infty}dx'\phi(x').
\end{equation}
The Sobolev optical depth $\tau_\mu$ is given by
(Castor \cite{cassob}, Rybicki \& Hummer \cite{rybashumrem})
\begin{equation}
\label{tau}
\tau_\mu=\frac{\io\rho\kappa\vth r}{\xa\io\vr\zav{1+\sigma\mu^2}},
\end{equation}
where the variable $\sigma$ was introduced by Castor~(\cite{cassob})
\begin{equation}\label{sigma}
\sigma=\frac{d\ln \io\vr}{d\ln r} -1.
\end{equation}

Inserting the solution of the transfer equation (\ref{isob}) into the
expression for the GO heating Eq.(\ref{qdopob}) we obtain
\begin{multline}
\label{cakgojed}
\io Q^{\mathrm{GO}} =
\frac{2\pi\kappa\io\rho\vthi 
\tilde I_c
}{c}\int_{-1}^1d\mu
\hzav{D(\mu)-\frac{\beta_c}{\beta}} \times\\*\times
\int_{-\infty}^{\infty} dx\, x \phi(x) e^{-\tau_\mu\Phi(x)}.
\end{multline}
The effect of line ensemble is usually described using the concept of a
line-strength distribution function (CAK, Abbott \cite{abpar}, Puls et
al. \cite{rozdelfce})
\begin{equation}
\label{cakrozdel}
dN(\kappa)=-N_0
\zav{\frac{\el{\rho}/\zav{W \el{m}}}{10^{11}\mbox{cm}^{-3}}}^{\delta} 
\kappa^{\alpha-2}d\kappa\frac{d\nu}{\nu},
\end{equation}
where normalization constant $N_0$ is taken in the form of
\begin{equation}
\label{cakrozdelcarn0}
N_0=\frac{c k}{\vth}\zav{1-\alpha}\alpha\, \sigma_e^{1-\alpha}.
\end{equation}
The
GO
heating formula for this line ensemble can be obtained by the
integration of the
heating term for one line (Eq.\ref{cakgojed}) over
the CAK distribution function Eq.(\ref{cakrozdel}).
In this case
it
takes the form of 
\begin{multline}
\io Q^{\mathrm{GO}} = \frac{\io\rho\vth\vthi L N_0}{2\pi c^2\xa  R_*^2}
\zav{\frac{\el{\rho}/\zav{W \el{m}}}{10^{11}\mbox{cm}^{-3}}}^{\delta}
\int_0^{\infty}d\kappa\,\kappa^{\alpha-1} \times\\*\times
\int_{-1}^1d\mu \hzav{D(\mu)-\frac{\beta_c}{\beta}}
\int_{-\infty}^{\infty} dx\, x \phi(x) e^{-\tau_\mu\Phi(x)},
\end{multline}
where $
\tilde I_c
=\DeltanuD L/4\pi^2 R_*^2$ was used.
Finally, applying substitution $y=\kappa\io\rho\vth r/\xa\io\vr$
preceding equation becomes
\begin{multline}
\label{dopa}
\io Q^{\mathrm{GO}} = \frac{\io\rho\vthi k \sigma_e L \zav{1-\alpha}\alpha}{2\pi c \xa R_*^2}
\zav{\frac{\el{\rho}/\zav{W \el{m}}}{10^{11}\mbox{cm}^{-3}}}^{\delta} \times\\*\times
\zav{\frac{\xa\io\vr}{\sigma_e\io\rho\vth r}}^{\alpha} G(\sigma,\mu_*),
\end{multline}
where the function $G(\sigma,\mu_*)$ is given by the triple integration
\begin{multline}
\label{defg}
G(\sigma,\mu_*)=\int_0^{\infty}dy\,y^{\alpha-1}\int_{-1}^1d\mu
\hzav{D(\mu)-\frac{\beta_c}{\beta}} \times\\*\times
\int_{-\infty}^{\infty} dx\, x \phi(x) 
\exp\zav{-\frac{y\Phi(x)}{{1+\sigma\mu^2}}}
\end{multline}
and in the integrals for $\beta_c$ and $\beta$ Eqs.~(\ref{betac}, \ref{betac}) 
the Sobolev depth shall be computed using
\begin{equation}
\tau_\mu=\frac{y}{1+\sigma\mu^2}
\end{equation}
instead of Eq.~(\ref{tau}).
Contrary to the radiative force formula (Castor \cite{cassob}) the
GO heating formula depends on the absorption profile.
For the determination of the
GO
heating term we selected Gaussian profile
(which comes from the Maxwellian velocity distribution).

\section{Model equations}\label{ModEq}

\subsection{Basic assumptions}

We assume that the stationary, spherically symmetric stellar wind
consists of three components, namely absorbing ions, nonabsorbing
hydrogen atoms and ions,
and electrons, denoted
by subscripts $\mathrm{i}$, $\mathrm{p}$, $\mathrm{e}$, respectively.
Each of them is described by a density $\rho_a$, radial velocity
${\vr}_a$, temperature $T_a$, electrical charge $q_a=e z_a$
(where $e$ is an elementary charge
and $z_a$ denotes the ionization degree
-- may have a  non-integer value),
and particle mass $m_a$.
Subscript $a$ stands for $a=\mathrm{i,p,e}$.
Contrary to our previous models (KKI), we allow for different
temperature of each component and for radial changes of electrical
charge.
We assume that chemical composition is given by the factor $z_*$, which
is a stellar metallicity relative to the solar value.

\subsection{Continuity equations}

In the case of a stationary spherically symmetric stellar wind each
component is described by the continuity equation Eq.(\ref{rovmomkont})
 in the form of
\begin{subequations}
\begin{equation}
\label{kontrov}
\frac{1}{r^2}\frac{d}{d r}\zav{r^2\rho_a{\vr}_a}  =  S_a,
\end{equation}
where the term $S_a$ accounts for radial change of mass-loss rate of
individual components due to the ionization.
Whereas for
all
types of ions the mass-loss rate is constant through the wind and thus
number of these particles is conserved ($\io S=\pr S =0$), 
we account for the possibility of variation of electron number via
ionization and recombination.
Because the total electric charge is conserved,
$$\sum_a \frac{d}{d r}\zav{r^2q_a\frac{\rho_a}{m_a}{\vr}_a}=0$$
and continuity equation holds
separately
for all components except electrons, we
obtain electron continuity equation in the form of
\begin{equation}
\label{kontrovel}
\frac{1}{r^2}\frac{d}{d r}\zav{r^2\el\rho\el\vr}=
\sum_{a\neq\mathrm{e}}\frac{\el m}{m_a}
 \rho_a{\vr}_a \frac{d z_a}{dr}.
\end{equation}
\end{subequations}

Although inclusion of a term $\el S$ into the electron continuity
equation does not significantly alter the model, it is important to
obtain well converged model.

\subsection{Momentum equations}

In the case of stationary spherically symmetric stellar wind the momentum
equation Eq.(\ref{rovmompohc}) has the form of
\begin{multline}
\label{rovhyb}
{\vr}_a\frac{d {\vr}_a}{dr}=
{g}_{a}^{\mathrm{rad}}-g-\frac{1}{{\rho}_a}\frac{d}{dr}\zav{{a}_a^2{\rho}_a}
 +\frac{q_a}{m_a}E+ \\*
                +\frac{1}{{\rho}_a}
\sum_{b\neq a} K_{ab}G(x_{ab})\frac{{\vr}_b-{\vr}_a}{|{\vr}_b-{\vr}_a|},
\end{multline}
where square of isothermal sound speed  is ${a}_a^2=kT_a/{m}_a$,
$E$ is a charge separation electric field.
Gravitational acceleration has the form $g={G{\eu M}}/{r^2}$,
where $\eu M$ is the stellar mass and $G$ is the gravitational constant.
The radiative acceleration acting on free electrons can be expressed as
\begin{equation}
\el{g}^{\mathrm{rad}}=\frac{\pr m}{\el m}\frac{G\Gamma{\eu M}}{r^2},
\end{equation}
where the ratio of the radiative force caused by absorption of
radiation by free electrons and gravitational force is
\begin{displaymath}
\Gamma=\frac{\sigma_e L}{4\pi c G{\eu M}}
\end{displaymath}
$L$ is the stellar luminosity and $\sigma_e$ is 
the mass scattering coefficient of the free electrons
(do not confuse it with Thomson scattering cross section).

The radiative acceleration acting on absorbing ions is taken in the
form of Castor et al. (\cite{cak})
\begin{equation}
\label{zarzrych}
\io{g}^{\mathrm{rad}}= \frac{1}{\xa} \frac{\sigma_e L}{4\pi r^2 c} f
     \zav{\frac{\el{\rho}/\zav{W \el{m}}}{10^{11}\mbox{cm}^{-3}}}^{\delta}
       k\zav{\frac{\xa}{\sigma_e\vth\io{\rho}} \frac{d\io{\vr}}{dr} }^{\alpha},
\end{equation}
with force multipliers $k$, $\alpha$, $\delta$ after Abbott
(\cite{abpar}).
The finite disk correction factor (Friend \& Abbott \cite{fa},
Pauldrach et al. \cite{ppk}) is
\begin{equation}
f=\frac{\zav{1+\sigma}^{\alpha+1}-
       \zav{1+\sigma\mu_c^2}^{\alpha+1}}
      {\zav{\alpha+1}\zav{1-\mu_c^2}\sigma\zav{1+\sigma}^{\alpha}},
\end{equation}
where 
$W$ is a dilution factor.
The thermal speed $\vth=\sqrt{2k\io T/\pr m}$ is, owing to the Abbott's
normalization, hydrogen thermal speed.
However, because $\vth$ physically describes ionic thermal speed, it
depends on ionic temperature.
Finally, $\xa$ is the photospheric ratio of the metallic ion density to
the passive plasma density.
We selected the same value as in KKI, namely $\xa=0.0127$ 
\zm{(this value corresponds to the solar ratio of sum of densities of
C, N, O, Fe to the density of bulk plasma)}.

Constant of friction evaluated using Fokker-Planck approximation (cf.
Burgers \cite{burgers}) has the following form:
\begin{equation}
{K}_{ab}={n}_a{n}_b\frac{4\pi  {q}_a^2{q}_b^2}{k T_{ab}}\ln\Lambda,
\end{equation} 
where ${n}_a$ and ${n}_b$ are number densities of individual components
and mean temperature of both components
\begin{equation}
T_{ab}=\frac{m_aT_b+m_bT_a}{m_a+m_b}.
\end{equation}
The Coulomb logarithm is of the form
\begin{equation}
\ln\Lambda=
\ln\hzav{\frac{3k\el T}{e^2}\zav{\frac{k\el T}{4\pi n e^2}}^{1/2}},
\end{equation}
where $n$ is the particle density ($n=\pr{n}+\el{n}+\io{n}$).
Finally, the Chandrasekhar function $G(x)$, defined in terms
of the error function $\erf(x)$ (Dreicer \cite{dreicer1}) is
\begin{equation}
G(x)=\frac{1}{2 x^2}\zav{\erf(x)-\frac{2x}{\sqrt\pi}\exp\zav{-x^2}}.
\end{equation}
The argument of the Chandrasekhar function is
\begin{equation}
x_{ab}=\frac{|{\vr}_b-{\vr}_a|}{\alpha_{ab}},
\end{equation}
where
\begin{equation}
\alpha_{ab}^2=\frac{2k\zav{m_aT_b+m_bT_b}} {m_am_b}.
\end{equation}

\subsection{Energy equation}

Energy equation (\ref{rovtepbur}) in the case of a stationary,
spherically symmetric multicomponent flow has the form of (cf. Burgers
\cite{burgers})
\begin{multline}
\label{rovenerg}
\frac{3}{2}k{\vr}_a\frac{\rho_a}{m_a}\frac{dT_a}{dr}+
a_a^2\rho_a
\frac{1}{r^2}\frac{d}{dr}\zav{r^2 {\vr}_a}=
Q_a^{\mathrm{rad}}+\\*+
\frac{1}{\sqrt\pi}\sum_{b\neq a}K_{ab} \frac{2k\zav{T_b-T_a}}{m_a+m_b}
 \frac{\exp\zav{-x_{ab}^2}}{\alpha_{ab}}+\\*+
 \sum_{b\neq a}\frac{m_b}{m_a+m_b}K_{ab}G(x_{ab})|{\vr}_b-{\vr}_a|.
\end{multline}
Two terms on the left-hand side stand for advection
and adiabatic cooling, respectively.
The right-hand side terms describe radiative heating/cooling, heat
exchange by encounters of different particles caused by unequal
temperatures of the components, and frictional heating.

There are two sources of radiative heating/cooling.
First source are bound-free and free-free transitions and the second is
\GO\ heating/cooling.

\subsubsection{Radiative energy term of bound-free and free-free
transitions}

Bound-free and free-free transitions (which will be called ``classical''
radiative transitions) deposits energy directly to electrons.
Therefore, this classical radiative energy term should be considered in
electron energy equation.
We decided to estimate the radiative heating/cooling term
$Q^{\mathrm{rad}}$ using two mechanisms only, hydrogen Lyman
bound-free and free-free transitions.
The detailed form of heating and cooling in the above mentioned
transitions is nearly the same as in KKI and will not be repeated here
(see also Kub\' at et al. \cite{kubel}).
The only difference is that the temperature in these equations is now
the electron temperature.
$J_{\nu}$ at the base of the wind is taken as an emergent
radiation from a spherically symmetric static hydrogen model
atmosphere for a corresponding stellar type (Kub\'at \cite{M4}).

\subsubsection{\GO\ heating/cooling}

Contrary to bound-free and free-free transitions \GO\ heating/cooling
deposits energy directly to absorbing ions.
GO heating/cooling term has the following form (GO, Eq.(\ref{dopa}))
\begin{multline}
\label{qdop}
\io Q^{\mathrm{rad}}\equiv\io Q^{\mathrm{GO}}
=\io\rho\io{\dot{q}}^{\mathrm{GO}}=
 \frac{\io\rho\vthi k \sigma_e L \zav{1-\alpha}\alpha}{2\pi c \xa R_*^2}
\times
\\*\times
\zav{\frac{\el{\rho}/\zav{W \el{m}}}{10^{11}\mbox{cm}^{-3}}}^{\delta}
\zav{\frac{\xa\vr}{\sigma_e\io\rho\vth r}}^{\alpha} G(\sigma,\mu_*),
\end{multline}
where the function $G(\sigma,\mu_*)$ is given by Eq.(\ref{defg}).
Here $\vthi$ is a thermal speed of driving ions, thus,
$\vthi=\sqrt{2k\io T/\io m}$. 
Inspecting the GO heating term (\ref{qdop}) we conclude that wind
is heated via the Doppler effect by direct radiation whereas the \GO\
cooling is caused by the diffuse radiation.
The sign of GO heating depends on the value of a variable $\sigma$.
In the case where $\sigma=0$, when the expansion is locally isotropic,
the GO heating and cooling vanishes.
When the $\sigma$ is positive, the cooling term dominates and the wind
is cooled by the GO heating.
In the opposite case, when the $\sigma$ is negative, the heating term
dominates and thus, the wind is heated.

\subsection{Charge separation electric field}

The equation for charge separation electric field can be obtained
directly from the third Maxwell equation, which in the case of spherical
symmetry can be written as
\begin{equation}
\label{erov}
\frac{1}{r^2}\frac{d}{dr}\zav{r^2 E}
=4\pi\sum_aq_an_a.
\end{equation}

\subsection{Ionic charge}

The ionization structure of stellar wind should be derived using time
consuming NLTE calculations (e.g. Pauldrach et al. \cite{modelydva}).
Because we want to determine only a mean charge of selected elements,
we use simpler approximate method.
As described by Mihalas (\cite{mih}, Eq. (5.46) therein), the ionization
equilibrium in stellar winds can be approximated by 
\begin{multline}
\label{ionrov}
\frac{n_{a,j}}{n_{a,j+1}}\approx \frac{1}{2} \frac{\el n}{W}
 \frac{U_{a,j}}{U_{a,j+1}}
 \zav{\frac{h^2}{2\pi m k T_R}}^{3/2}
 \zav{\frac{T_R}{\el T}}^{1/2}\times\\*
 \times\exp\zav{\frac{\chi_{a,j}}{k T_R}},
\end{multline}
where $\chi_{a,j}$ is the ionization potential,
$U_{a,j}$ is the partition function and
$T_R$ is the radiation temperature (we
set
$T_R=\frac{3}{4}\Teff$).
Partition function approximations are taken from Smith \& Dworetsky
(\cite{smid}).
Electrical charge of absorbing and passive ions are then derived using
a formula
\begin{equation}
z_a=\frac{\sum_j j n_{a,j}}{\sum_jn_{a,j}}.
\end{equation}
Clearly, electron electrical charge is $\el z=-1$ everywhere. 

\subsection{Critical points}
   
Critical points are points where derivatives of variables cannot be
determined directly from differential equations.
The derivation of critical point conditions for our set of equations is
simpler than that of KKI because we use different temperatures of each
component and correct momentum equation for electrons.
Due to these
generalizations
the set of critical point equations is not as coupled as it was in KKI
and thus, the obtained critical point conditions are simpler.

We write model equations in a simplified form, where we explicitly write
only terms containing derivatives of individual variables and other
terms are included into the terms $F_i$.
Thus, the continuity equations~(\ref{kontrov},\ref{kontrovel}) are
\begin{subequations}
\label{kritsyst}
\begin{equation}
\label{krithust}
\rho_a\frac{d{\vr}_a}{d r}+{\vr}_a\frac{d\rho_a}{d r}=
F_{a,1}(r,\rho_a,{\vr}_a).
\end{equation}
In the electron continuity equation (\ref{kontrovel}) we neglected the
derivatives of ionic charge because their contribution to electron
continuity equation is only marginal.
However, inclusion of such term influences critical point and regularity
conditions for electrons only, which will not be used (see bellow).

Similarly we can rewrite momentum equations~(\ref{rovhyb}).
In the momentum equations of absorbing ions we shall linearize
a term containing the velocity gradient.
Note that because model equations are not quasi-linear (i.e. linear
with respect to the derivatives of the independent variables),
mathematically more correct method would employ some form of
transformation to the quasi-linear form (cf. Courant \& Hilbert
\cite{drina}).
However, because the results are essentially the same in this case,
we present analysis of critical points in a simplified form.
Thus, momentum equations are
\begin{multline}
\label{krithyb}
{\vr}_a\frac{d {\vr}_a}{dr}-
\frac{\partial {g}_{a}^{\mathrm{rad}}}{\partial\zav{d\vr_a/dr}}
\frac{d {\vr}_a}{dr}+
\frac{{a}_a^2}{{\rho}_a}\frac{d{\rho}_a}{dr}+
\frac{{a}_a^2}{T_a}\frac{dT_a}{dr}=
\\*= F_{a,2}(r,E,\rho_b,\vr_b,T_b,z_b).
\end{multline}
Similarly, due to the dependence of the Doppler term on the velocity
gradient (in the Sobolev approximation) we shall write energy equations
in the form of
\begin{multline}
\label{krittep}
\frac{3}{2}{\vr}_a\rho_a\frac{{a}_a^2}{T_a}\frac{dT_a}{dr}+
a_a^2\rho_a \frac{d{\vr}_a}{dr}-
\rho_a\frac{\partial\dot{q}_a^{\mathrm{rad}}}{\partial\zav{d\vr_a/dr}}
\frac{d {\vr}_a}{dr}=
\\*= F_{a,3}(r,\rho_b,\vr_b,T_b,z_b).
\end{multline}
The system of equations is closed by the equation for charge separation
electric field, which has a simple form,
\begin{equation}
\label{kritel}
\frac{dE}{dr}=F_{4}(r,\rho_b,z_b).
\end{equation}
\end{subequations}

The system of equations~(\ref{kritsyst}) can be simplified by inserting
the derivatives of density from the Eq.(\ref{krithust}) and derivatives
of temperature~(\ref{krittep}) into the momentum Eq.(\ref{krithyb}).
We obtain modified linearized
momentum equations
\begin{multline}
\label{krithybz}
\hzav{{\vr}_a-
\frac{\partial {g}_{a}^{\mathrm{rad}}}{\partial\zav{d\vr_a/dr}}-
\frac{5}{3}\frac{{a}_a^2}{{\vr}_a}+
\frac{2}{3\vr_a}\frac{\partial\dot{q}_a^{\mathrm{rad}}}
{\partial\zav{d\vr_a/dr}}}
\times\\*\times\frac{d {\vr}_a}{dr}=
 F_{a}(r,E,\rho_b,\vr_b,T_b,z_b),
\end{multline}
where $F_a$ is a combination of $F_{a,i}$ from Eqs. (\ref{krithust}),
(\ref{krithyb}), and (\ref{krittep}).
Because the charge separation electric field equation~(\ref{kritel})
does not introduce any physically interesting critical point,
the system of equations (\ref{krithybz}) consists of only three
independent critical point equations.
Each of them will be discussed separately in the following subsections.

\subsubsection{Critical point of passive ions}

For the passive plasma the critical point condition~(\ref{krithybz})
has a simple form 
\begin{equation}
\label{kritp}
\pr{v_r^2}=\frac{5}{3}\pr a^2.
\end{equation}
In order to obtain continuous solution of model equations,
at this point
should the quantity $F_{a}(r,E,\rho_b,\vr_b,T_b,z_b)$ vanish
(regularity condition). 
This requirement is fulfilled if
\begin{multline}
\label{regp}
\frac{10\pr{a}^2}{3r}-g+\frac{\pr q}{\pr m}E+
\frac{1}{\pr{\rho}}
\sum_{b\neq \mathrm{p}} K_{\mathrm{p}b}G(x_{\mathrm{p}b})
\frac{{\vr}_b-{\vr}_{\mathrm{p}}}{|{\vr}_b-{\vr}_{\mathrm{p}}|}-\\*
-\frac{2}{3}\frac{1}{\pr\rho\pr\vr}
\frac{1}{\sqrt\pi}\sum_{b\neq {\mathrm{p}}}
 K_{{\mathrm{p}}b} \frac{2k\zav{T_b-T_{\mathrm{p}}}}{m_{\mathrm{p}}+m_b}
 \frac{\exp\zav{-x_{{\mathrm{p}}b}^2}}{\alpha_{{\mathrm{p}}b}}- \\* -
 \frac{2}{3}\frac{1}{\pr\rho\pr\vr}
\sum_{b\neq {\mathrm{p}}}\frac{m_b}{m_{\mathrm{p}}+m_b}
K_{{\mathrm{p}}b}G(x_{{\mathrm{p}}b})|{\vr}_b-{\vr}_{\mathrm{p}}|=0.
\end{multline}
This equation is a generalization of well-known regularity condition
for the coronal wind.

\subsubsection{Critical point of absorbing ions}

Critical point condition Eq.(\ref{krithybz}) for absorbing ions has form
\begin{equation}
\io{\vr}-
\frac{\partial \io{g}^{\mathrm{rad}}}{\partial\zav{d\io\vr/dr}}-
\frac{5}{3}\frac{\io{a}^2}{\io{\vr}}+
\frac{2}{3\io\vr}
\frac{\partial\io{\dot{q}}^{\mathrm{rad}}}{\partial\zav{d\io\vr/dr}}=0.
\end{equation}
However, as was discussed by KKI, this critical point condition is not
met anywhere in the wind.
Therefore, to obtain CAK type solution of the stellar wind we use
another condition, which is connected to generalized Abbott waves
(see KKI).
Such condition can be derived by multiplying the critical point
conditions Eq.(\ref{krithybz}) by density of corresponding component and
summing them
\begin{multline}
\sum_a \rho_a \frac{d {\vr}_a}{dr} \hzav{{\vr}_a-
\frac{5}{3}\frac{{a}_a^2}{{\vr}_a}}+\\*+
\io\rho\frac{d\io\vr}{dr}\hzav{
\frac{2}{3\io\vr}
\frac{\partial\io{\dot{q}}^{\mathrm{rad}}}{\partial\zav{d\io\vr/dr}}-
\frac{\partial \io{g}^{\mathrm{rad}}}{\partial\zav{d\io\vr/dr}}}=\\*
= \sum_a F_{a}(r,E,\rho_b,\vr_b,T_b,z_b).
\end{multline}
To assure that common equation of motion (obtained by summing of
individual momentum equations~(\ref{rovhyb})) does not depend
on the derivatives of any variables, a condition
\begin{multline}
\label{cakpod}
\sum_a \rho_a \frac{d {\vr}_a}{dr} \hzav{{\vr}_a-
\frac{5}{3}\frac{{a}_a^2}{{\vr}_a}}+\\*+
\io\rho\frac{d\io\vr}{dr}\hzav{
\frac{2}{3\io\vr}
\frac{\partial\io{\dot{q}}^{\mathrm{rad}}}{\partial\zav{d\io\vr/dr}}-
\frac{\partial \io{g}^{\mathrm{rad}}}{\partial\zav{d\io\vr/dr}}}=0
\end{multline}
shall be fulfilled.
This condition we use to fix the mass-loss rate.

\subsubsection{Critical point of electrons}

The last critical point condition for electrons has again very simple
form
\begin{equation}
\label{krite}
\el{v_r^2}=\frac{5}{3}\el a^2.
\end{equation}
This critical point condition has similar form as the condition of 
nonabsorbing ions Eq.(\ref{kritp}).
Thus, regularity condition for electrons 
resembles the regularity condition for passive ions Eq.(\ref{regp}),
\begin{multline}
\label{rege}
\frac{10\el{a}^2}{3r}+\el{g}^{\mathrm{rad}}-g+\frac{\el q}{\el m}E+
\frac{1}{\el{\rho}}
\sum_{b\neq \mathrm{e}} K_{\mathrm{e}b}G(x_{\mathrm{e}b})
\frac{{\vr}_b-{\vr}_{\mathrm{e}}}{|{\vr}_b-{\vr}_{\mathrm{e}}|}-\\*
-\frac{2}{3}\frac{1}{\el\rho\el\vr}
\frac{1}{\sqrt\pi}\sum_{b\neq {\mathrm{e}}}
 K_{{\mathrm{e}}b} \frac{2k\zav{T_b-T_{\mathrm{e}}}}{m_{\mathrm{e}}+m_b}
 \frac{\exp\zav{-x_{{\mathrm{e}}b}^2}}{\alpha_{{\mathrm{e}}b}}- \\* - 
 \frac{2}{3}\frac{1}{\el\rho\el\vr} 
\sum_{b\neq {\mathrm{e}}}\frac{m_b}{m_{\mathrm{e}}+m_b}
K_{{\mathrm{e}}b}G(x_{{\mathrm{e}}b})|{\vr}_b-{\vr}_{\mathrm{e}}|- \\* -
\frac{2}{3}\frac{1}{\el\rho\el\vr}\el{Q}^{\mathrm{rad}}=0.
\end{multline}
However, numerical tests showed that electron regularity condition 
(\ref{rege}) is approximately fulfilled at the electron critical point
if the zero current condition is used as a boundary condition.
Thus, this condition should not be necessarily included into
the set of model equations.

\subsection{Boundary conditions}

\subsubsection{Boundary conditions for temperatures}

We assume that the flow at the inner boundary is in radiative
equilibrium and that the boundary temperature of all components is the
same, thus, we write boundary condition for temperatures in the form of
\begin{subequations}
\begin{gather}
\label{hrentepip}
T_{a}=\el T,\quad a=\mathrm{i, p},\\
Q^{\mathrm{rad}}(\rhvez)=0.
\end{gather}
\end{subequations}
Boundary values of ionic charges can be directly obtained from the
condition of ionization equilibrium (\ref{ionrov}).

\subsubsection{Boundary condition for velocity}

Conditions (\ref{regp}, \ref{cakpod}) can be generally used to fix the
boundary values of model quantities.
However, inclusion of two inner conditions directly into model equations
sometimes leads to
numerical problems.
Therefore, we use more secure method, which gives essentially the same
results.

We start to calculate our models at the passive plasma critical point.
Consequently, the boundary condition for the passive plasma velocity is
the critical point condition
Eq.(\ref{kritp}).
Boundary condition for
the velocity of absorbing ions
may be obtained from the passive plasma regularity condition
Eq.(\ref{regp}).
Because we suppose equal boundary temperatures of each component
Eq.(\ref{hrentepip}), the regularity condition may be simplified 
\begin{multline}
\label{hranp}
\frac{10\pr{a}^2}{3r}-g+\frac{\pr q}{\pr m}E+
\frac{1}{\pr{\rho}}
\sum_{b\neq \mathrm{p}} K_{\mathrm{p}b}G(x_{\mathrm{p}b})
\frac{{\vr}_b-{\vr}_{\mathrm{p}}}{|{\vr}_b-{\vr}_{\mathrm{p}}|}-\\*
- \frac{2}{3}\frac{1}{\pr\rho\pr\vr}
\sum_{b\neq {\mathrm{p}}}\frac{m_b}{m_{\mathrm{p}}+m_b}
K_{{\mathrm{p}}b}G(x_{{\mathrm{p}}b})|{\vr}_b-{\vr}_{\mathrm{p}}|=0.
\end{multline}

The boundary value of electron velocity
is chosen
to fulfil the electron regularity condition Eq.(\ref{rege}) at the
electron critical point Eq.~(\ref{krite}).
As was already mentioned, this condition is approximately satisfied
if the zero current condition
\begin{equation}
\label{elrych}
\el{n}\el{\vr}=\pr z\pr n\pr{\vr} + \io z \io n\io{\vr}
\end{equation}
is used as boundary condition for electron velocity.
The latter condition (\ref{elrych}) was applied in our models.

\subsubsection{Boundary conditions for density}

We write the boundary condition for the passive plasma density in the
same form as in	KKI, 
\begin{equation}
\label{hranio}
z_*\pr\rho(\rhvez)=
  \frac{1}{\xa} \io\rho(\rhvez) \frac{\io\vr(\rhvez)}{\pr\vr(\rhvez)}.
\end{equation}
Here we only newly introduced the relative abundance $z_*$ which
accounts for different chemical composition
in stellar atmospheres.

Boundary value of ionic density is determined
numerically to obtain CAK-type solution (see Sect.\ref{num}).
Boundary electron density is calculated from the condition of
quasi-neutrality
\begin{equation}
\label{elhust}
\el{n}=\sum_{a\neq\mathrm{e}}z_a n_a.
\end{equation}

\subsubsection{Boundary conditions for electric field}

Because we have not any critical point condition to determine the
intensity of the electric field at the stellar surface, we
used the condition of neutrality, which simply sets the gradient of
the electric field at the stellar surface to zero (cf. Eq.\ref{erov}).

\subsection{Numerical method}
\label{num}

We apply Henyey method (Henyey et al. \cite{heney}), which is a
modification of the well-known Newton-Raphson method to solve equations
described here together with the appropriate boundary conditions.
We use essentially the same method as KKI, except that the vector of
variables at each grid point~$d$ has the form of
\begin{equation}
\psi_d
=\zav{\rho_{a,d},v_{ra,d},T_{a,d},z_{a,d}, E_d,\Delta v_{r,d}},
\quad a=\mathrm{e},\mathrm{i},\mathrm{p}
\end{equation}
where the velocity difference
\begin{equation}
\Delta\vr=\io\vr-\pr\vr
\end{equation}
may be
added to the set of variables
to assure better convergence of the models.

First of all we search for the boundary density
$\rho_{0}=\io\rho(\rhvez)$.
We compute several wind models (each of them is a result of several
Newton-Raphson iterative steps) for the region near the star for
different values of $\rho_{0}$ (for more details see KK0, KKI).
We select such value of $\rho_{0}$ which allows wind model to pass
smoothly through the point defined by the Eq.(\ref{cakpod}) and to
obtain CAK-type solution.
After the appropriate value of $\rho_{0}$ is chosen, we compute wind 
model downstream the point defined by the Eq.(\ref{cakpod}) again using
several Newton-Raphson iterative steps.

Detailed method of calculation of 
\GO\
heating/cooling term is given
in Appendix~\ref{vypg}.

%%%%%%%%%%%%%%%%%%%%%%%%%%%%%%%%%%%%%%%%%%%%%%%%%%%%%%%%%%%%%%%%%%%%%%%%
%                     Results of calculations
%%%%%%%%%%%%%%%%%%%%%%%%%%%%%%%%%%%%%%%%%%%%%%%%%%%%%%%%%%%%%%%%%%%%%%%%

\section{Results of calculations}

\obrazekd{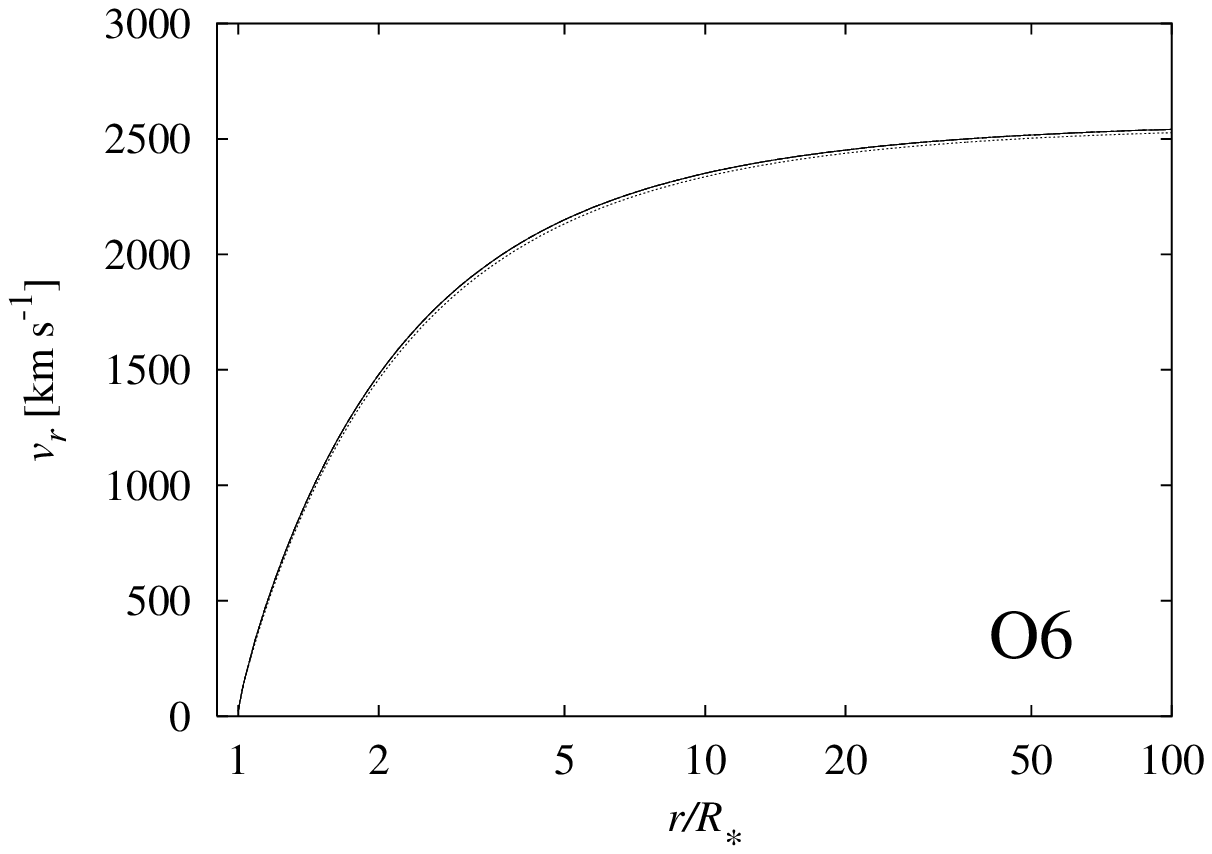}{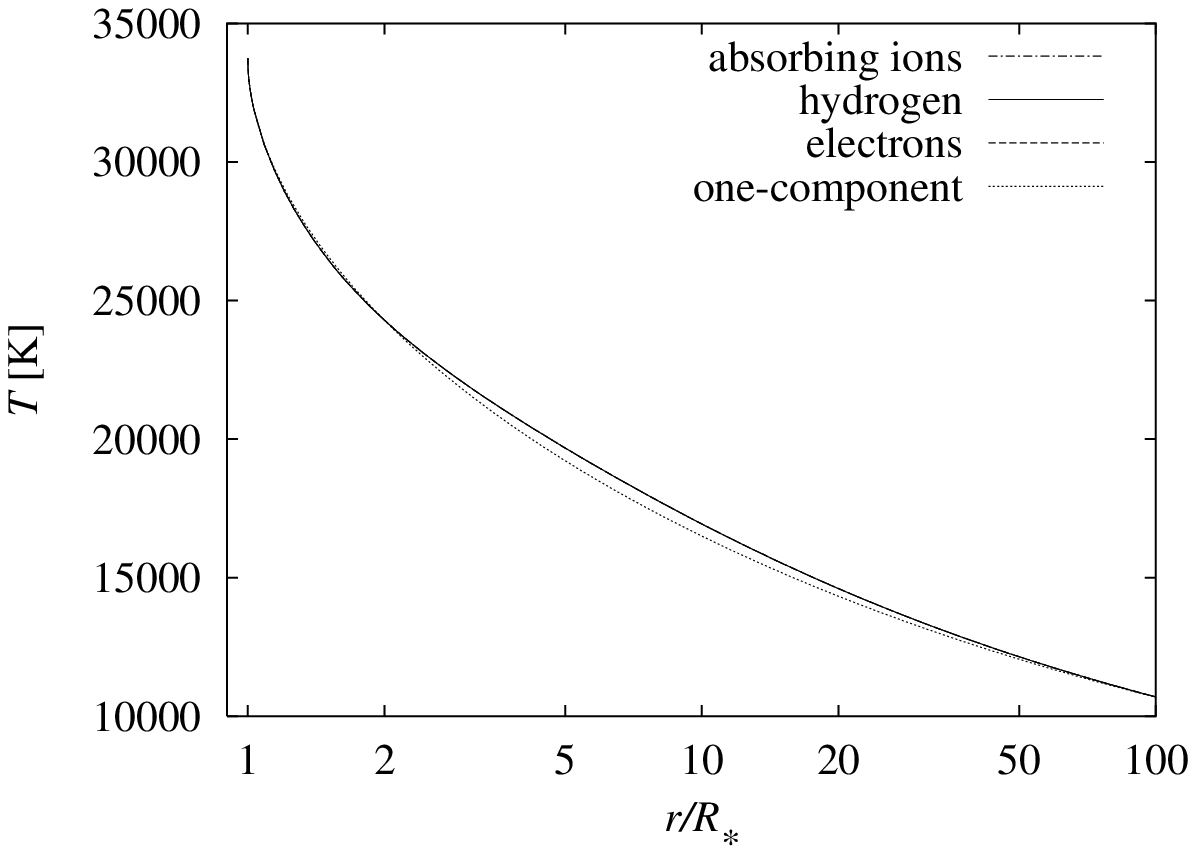}
{{\em Upper panel:} Comparison of the radial wind velocity of
  one-component (dotted line) and three-component radiatively driven
  stellar wind models of an 
\zm{O6}
  star.
  Radiatively accelerated ions are denoted using dashed-dotted line,
  passive plasma with full line and electrons with dashed line.
  Notice that all curves are very nearly the same.
{\em Lower panel:} Comparison of temperature stratification of
  one-component and three-component models.
  Assignment of all curves is the same as for velocity.
  Notice that the curves describing temperature of individual components
  of a three-component model coincide, which is not true if we compare
  one and three component models.
}
 {o6}

\obrazekd{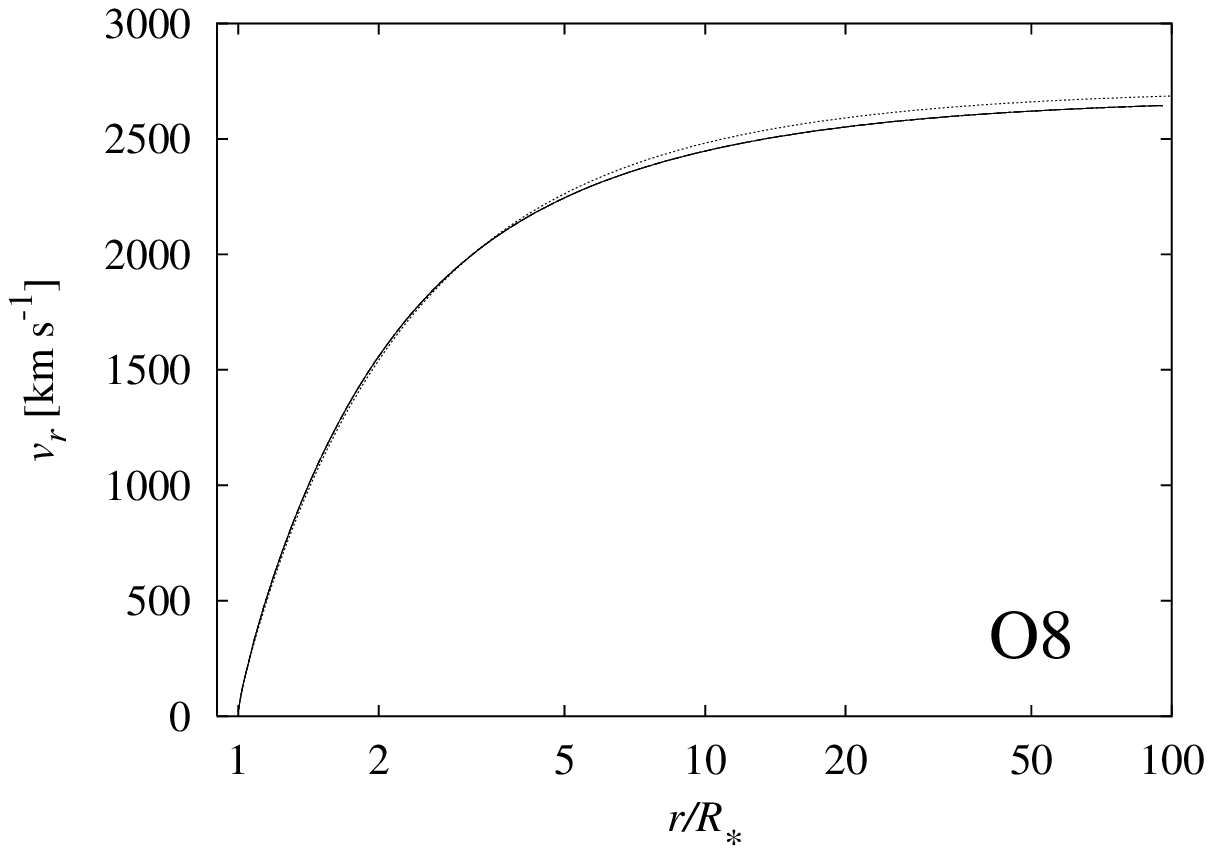}{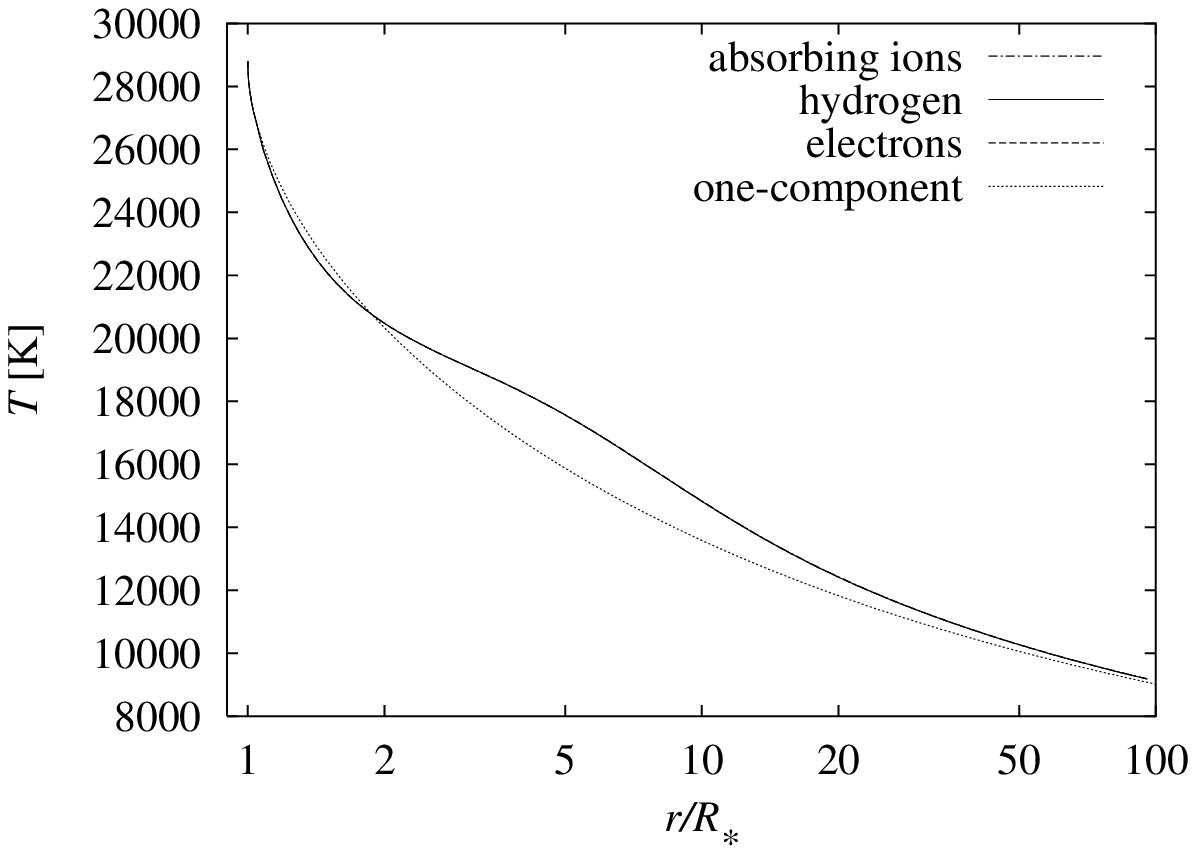}
{The same as Fig.\protect\ref{o6} for an O8 star.
 There is no significant difference in the temperatures of each
 component.
 The wind temperature of the one-component and three-component models
 differ due to the
 \GO\
 heating.
}{o8}

\obrazekd{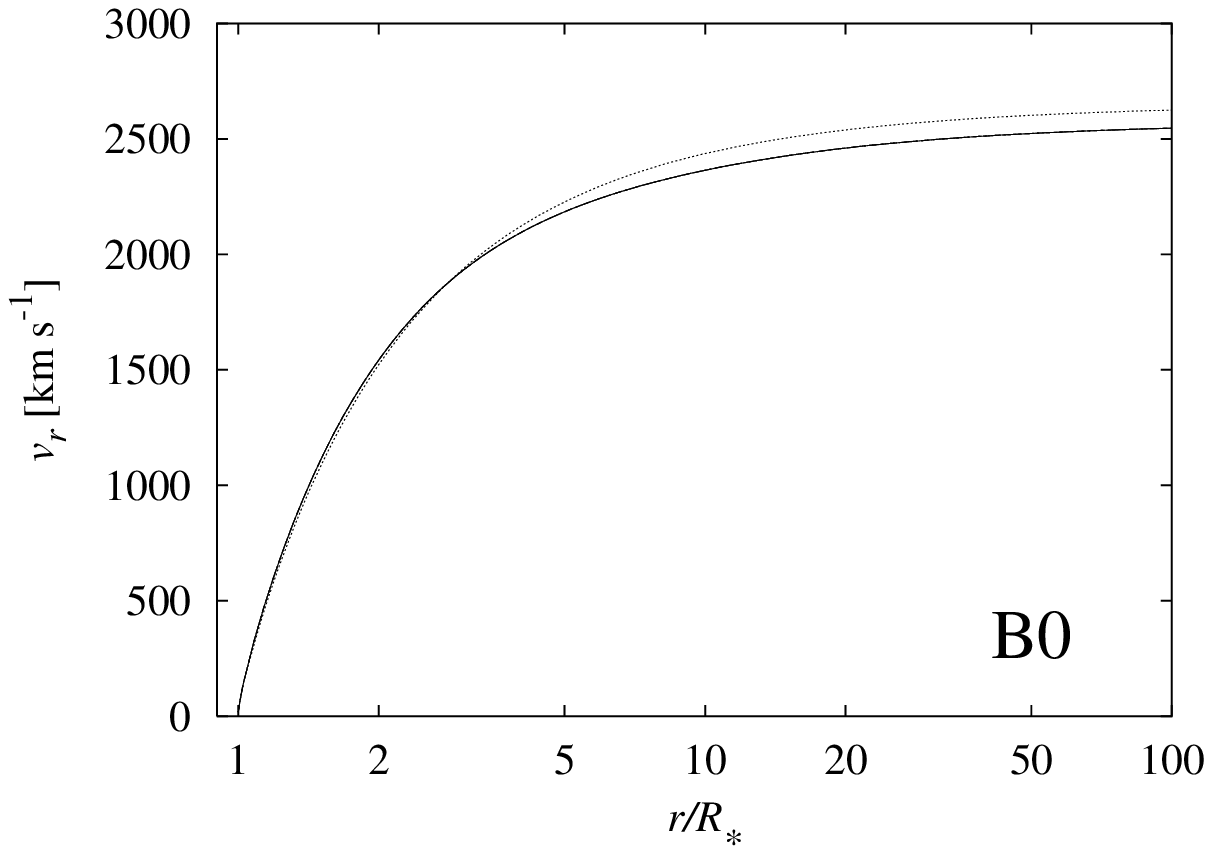}{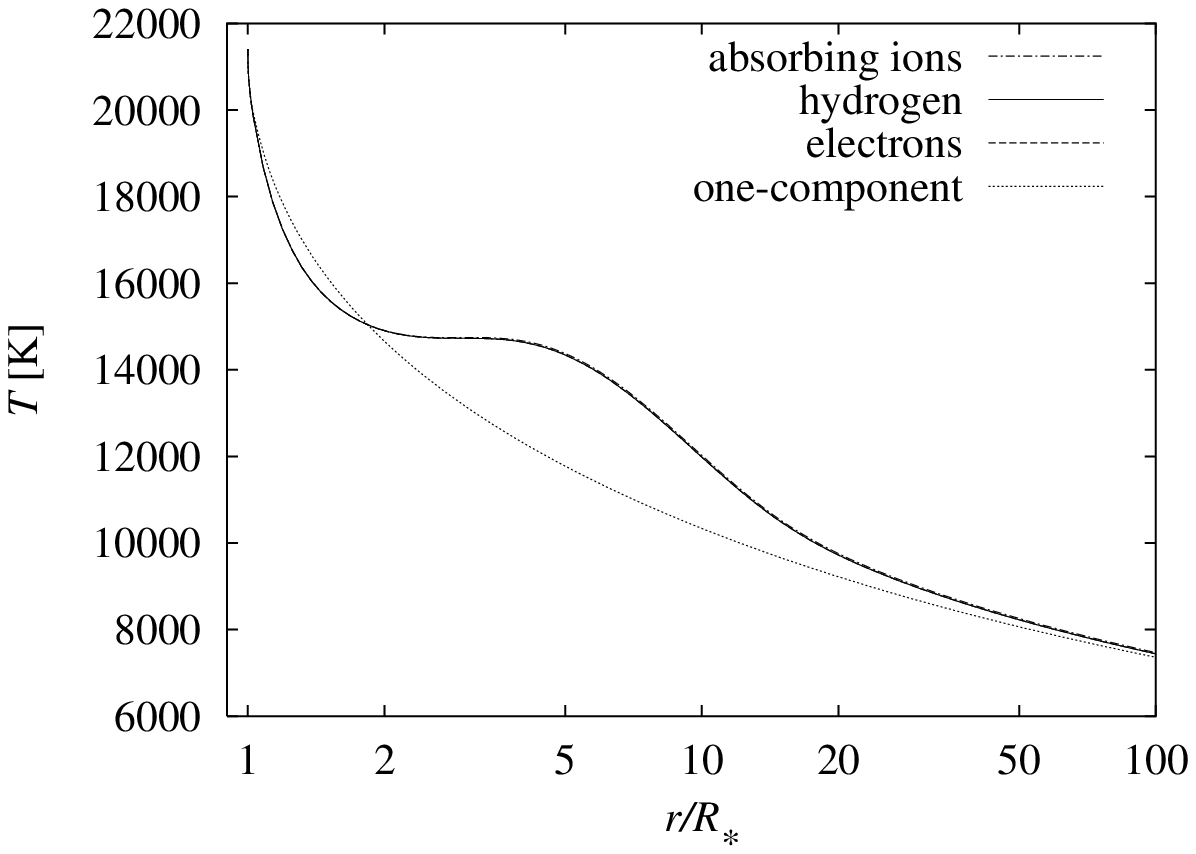}
{The same as Fig.\protect\ref{o6} for a B0 star.
 For this star 
 \GO\
 heating (in the outer parts of the wind) and
 cooling (in the inner parts of the wind) effects are important for
 temperature structure.
 Note that the frictional heating is negligible in this case
 and that the temperatures of particular components are nearly the same.
}
{b0}

\obrazekd{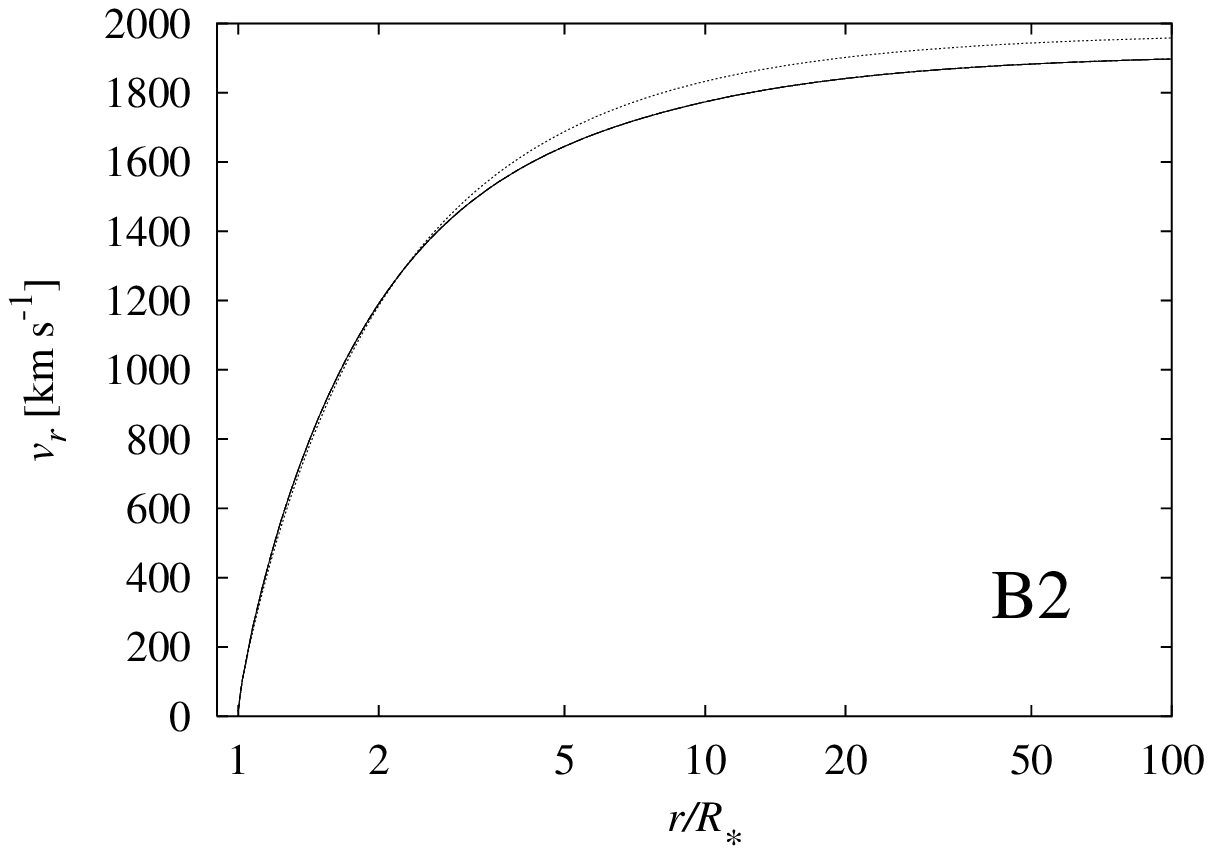}{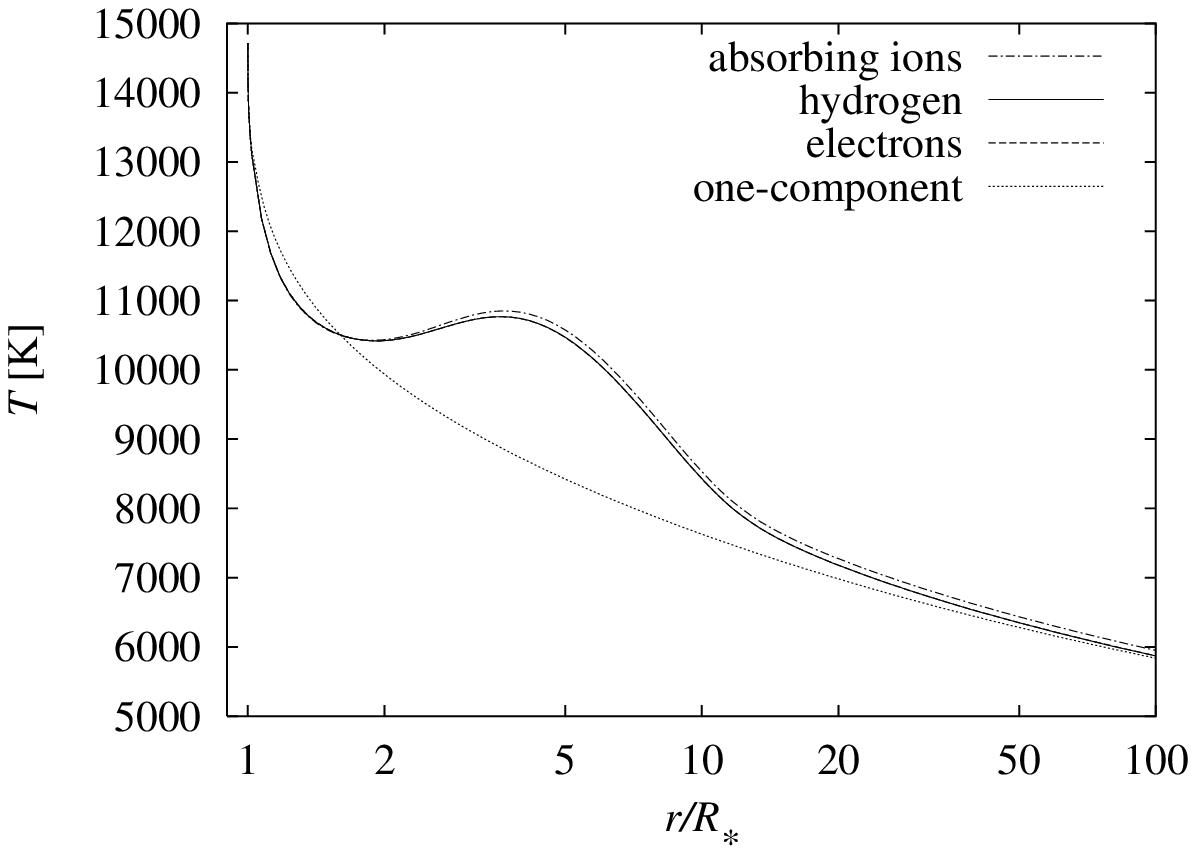}
{The same as Fig.\protect\ref{o6} for a B2 star.
 \GO\
 heating and cooling is greater than for a B0 star -- cf.
 Fig.\protect\ref{b0}.
 Frictional heating is negligible. 
 The temperatures of particular components are nearly the same.
}
{b2}

\obrazekd{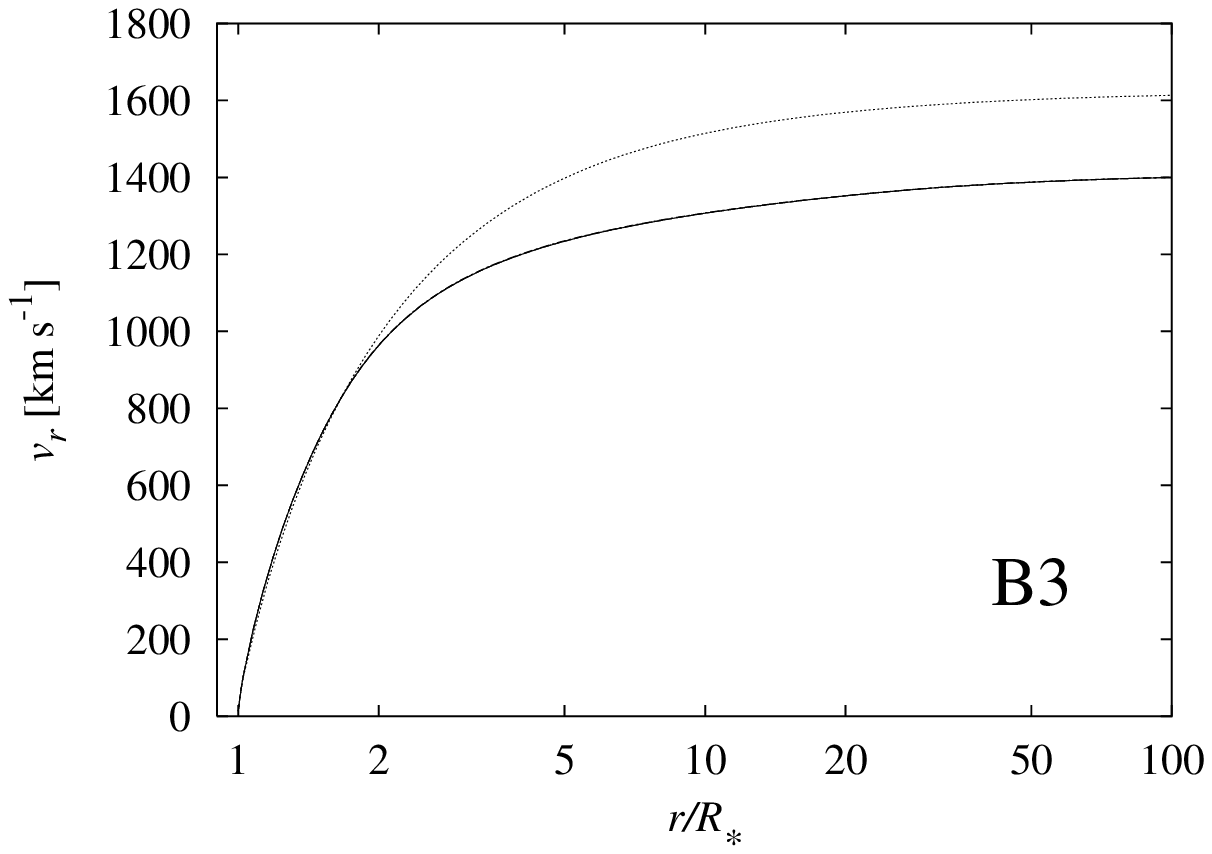}{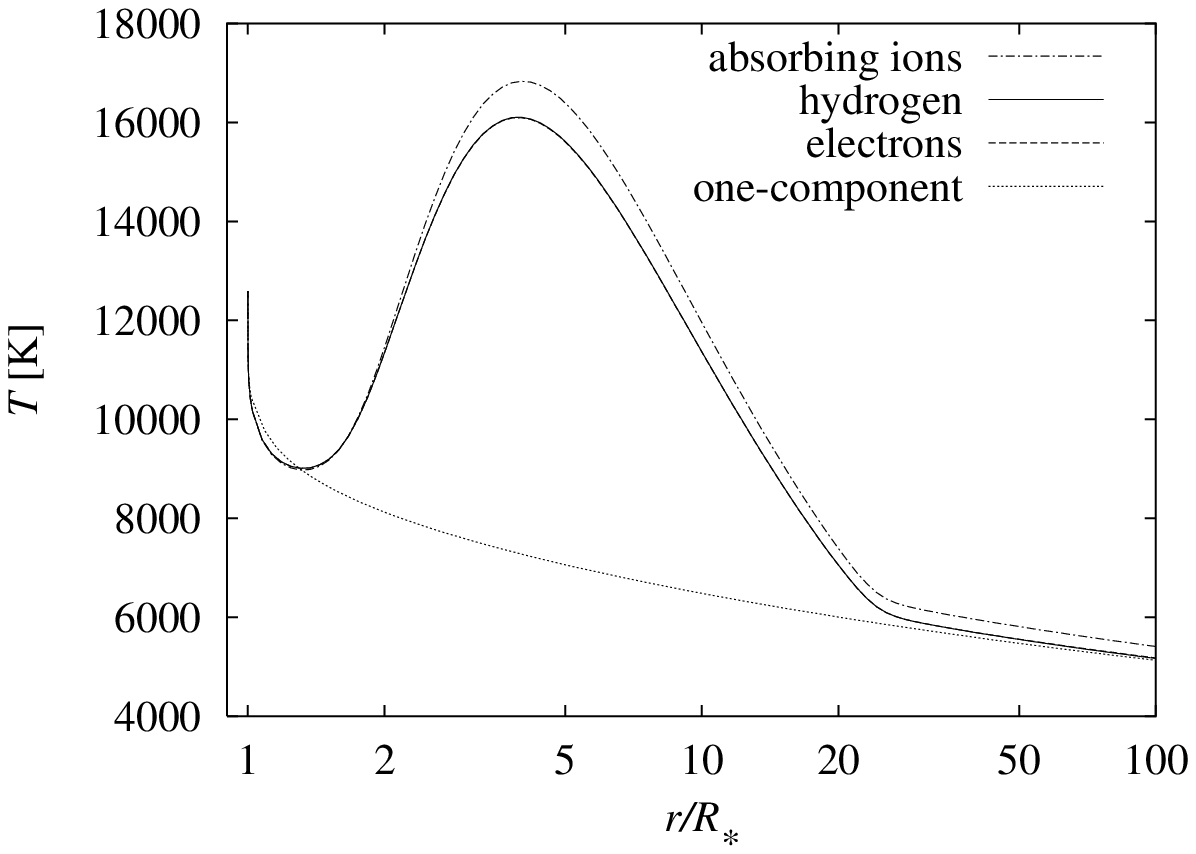}
{The same as Fig.\protect\ref{o6} for a B3 star.
The wind is heated by both frictional and 
\GO\
heating in the outer
parts of the wind.
The temperatures of absorbing ions and electrons are nearly the same
whereas the ionic temperature slightly differs mainly in the outer parts
of the wind.
}
{b3}

\obrazekd{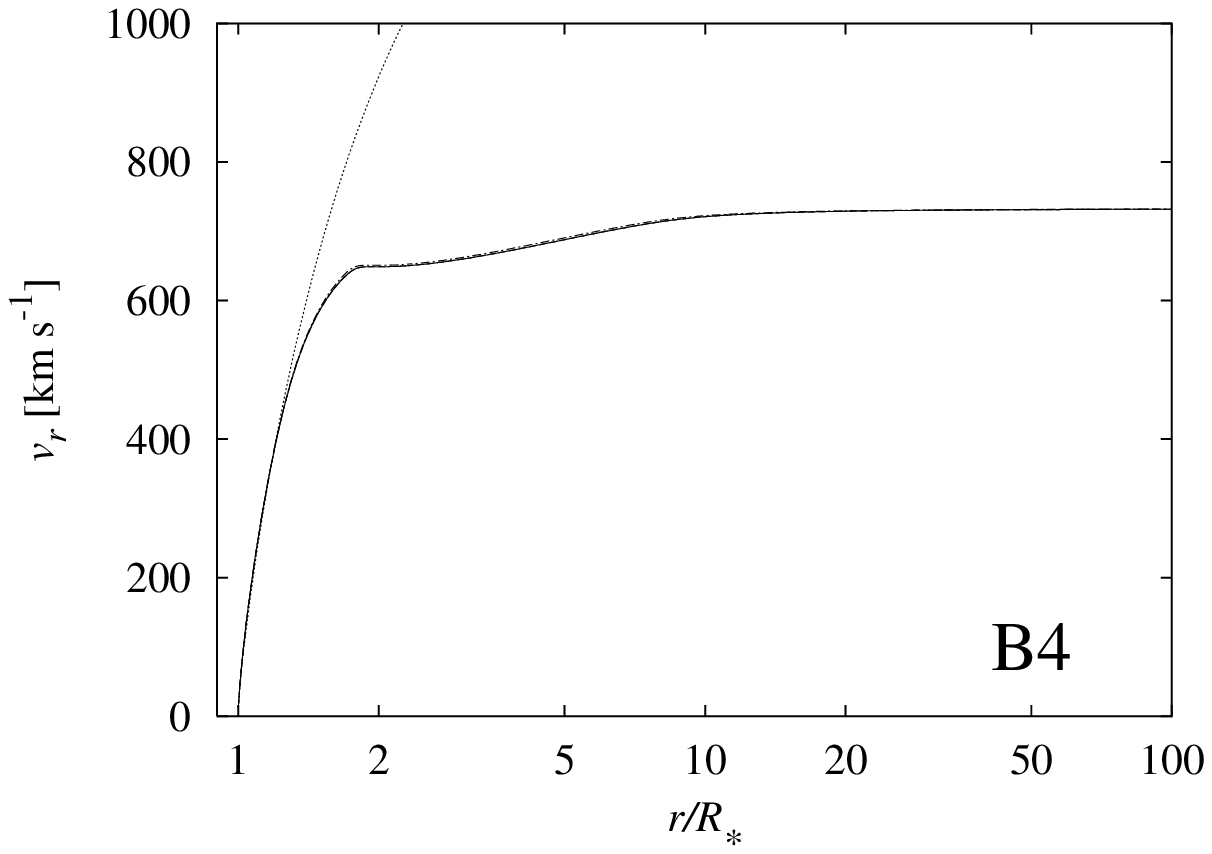}{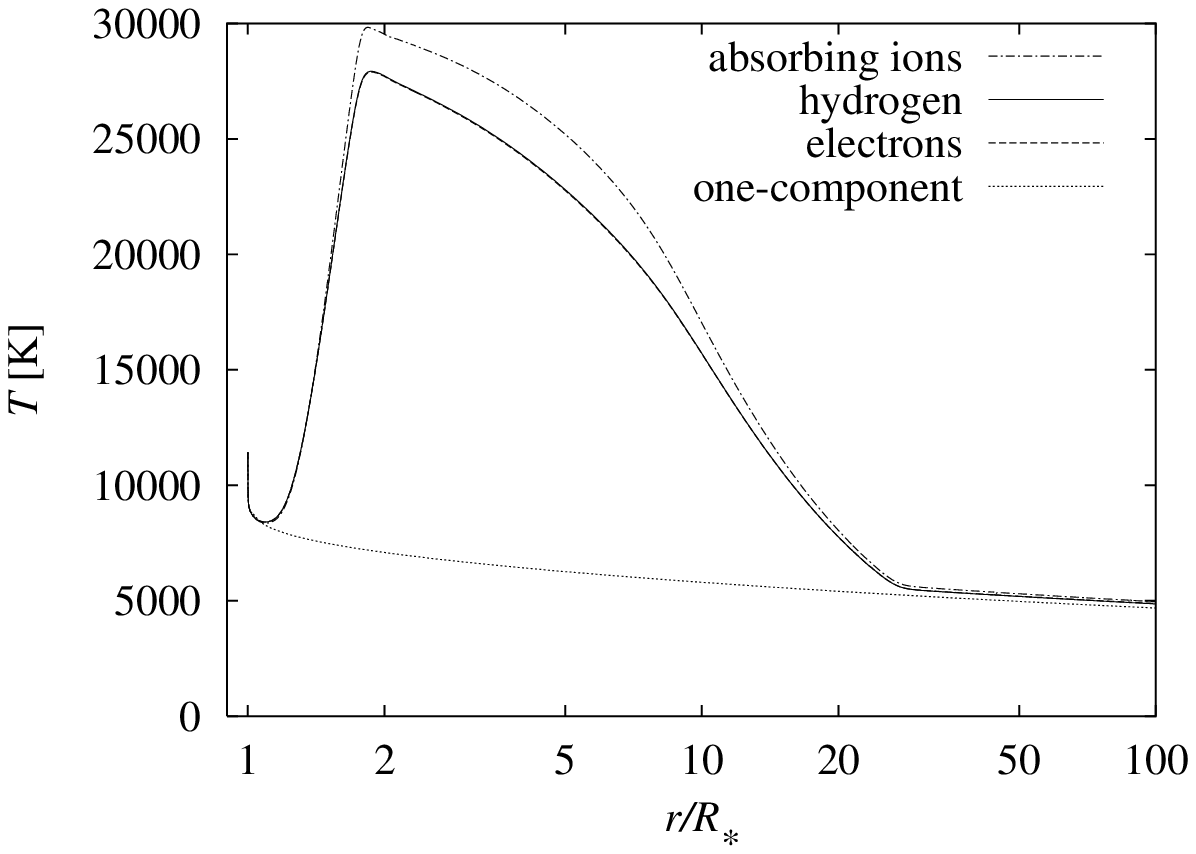}
{The same as Fig.\protect\ref{o6} for a B4 star.
The wind is heated by both frictional and 
\GO\
heating in the outer
parts of the wind.
Notice that the effect of heating is more pronounced than for the case
of a B3 star (Fig. \ref{b3}) and the temperature at its maximum is
larger than for a {\em cooler} star.
The temperatures of absorbing ions and electrons are nearly the same.
}
{b4}

\obrazek{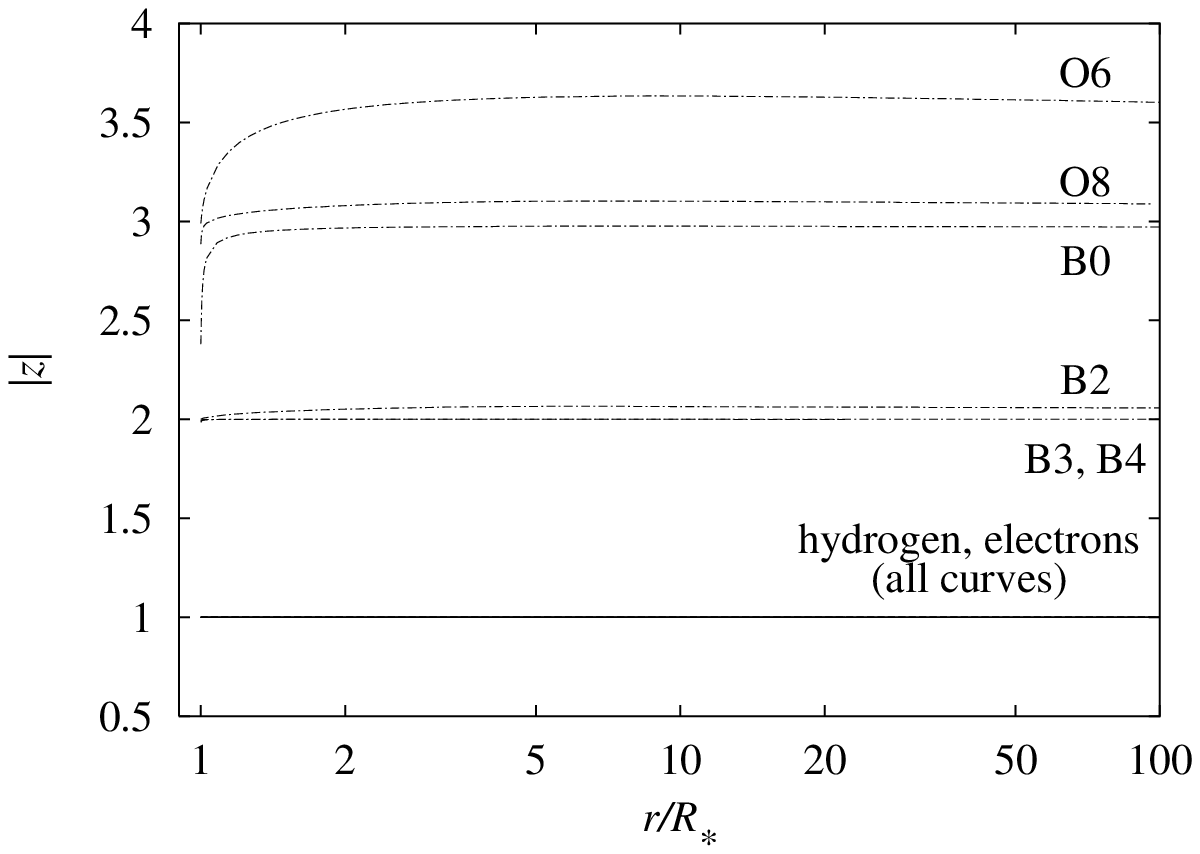}{The dependence of the ionic charge on radius for
different wind models.
Accelerated ions are denoted using dashed-dotted line.}{naboj}

\obrazekd{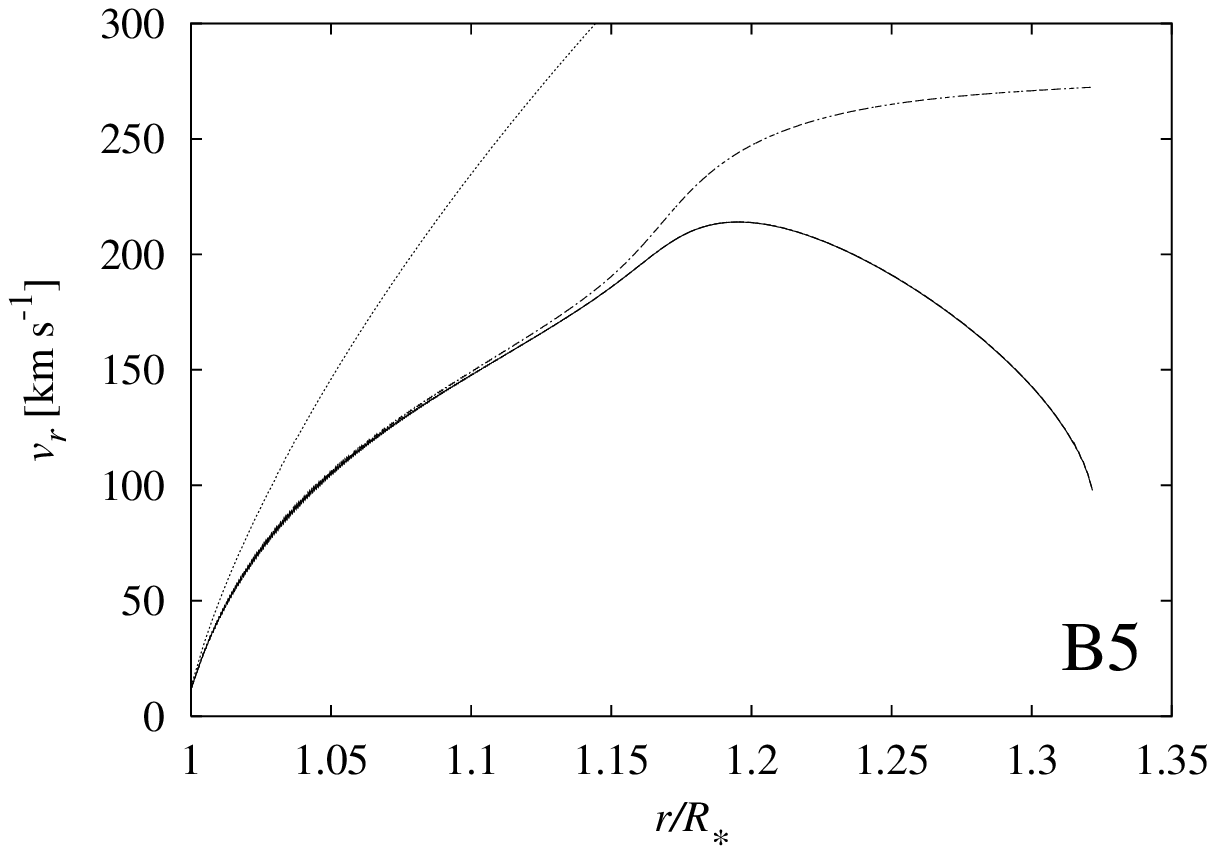}{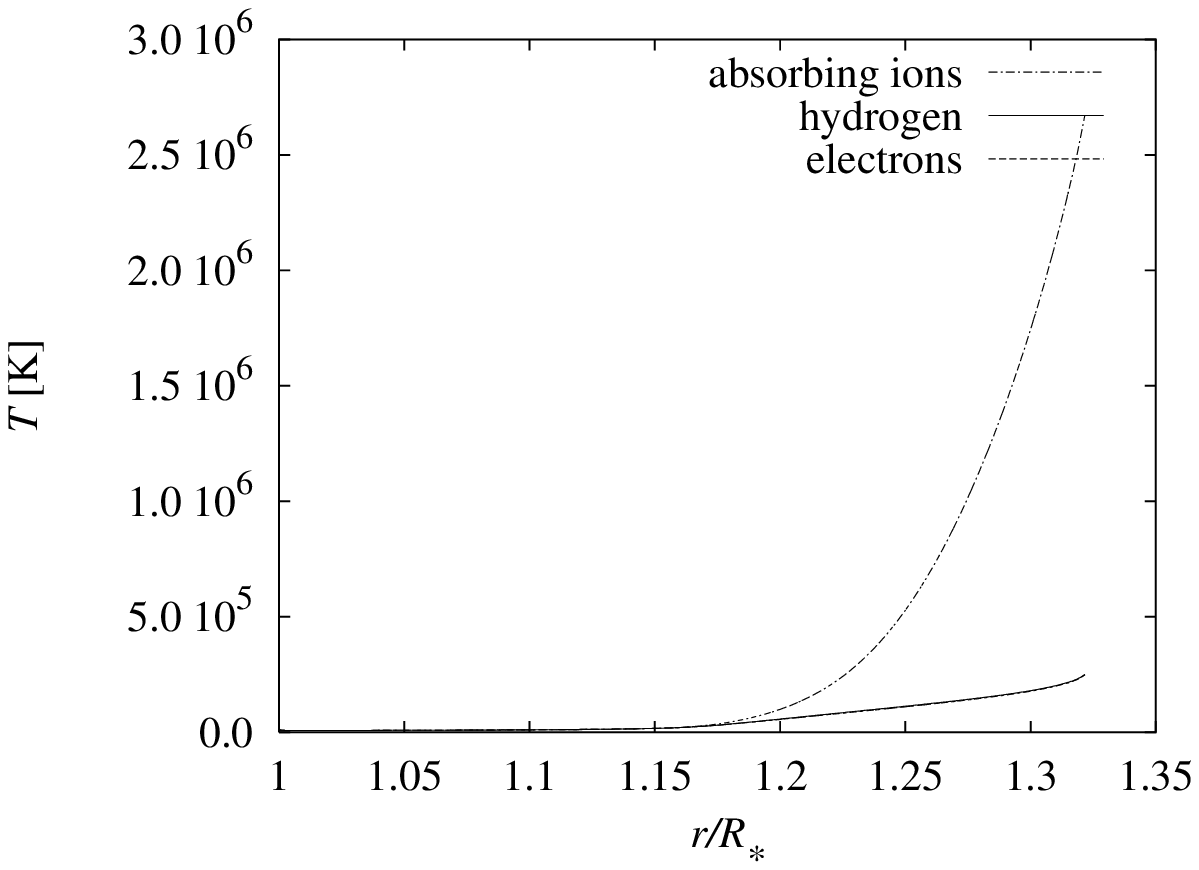}
{The same as Fig.\protect\ref{o6} for a B5 star.
The passive component decouples
just above the stellar surface
and subsequently falls back onto the star.
Note that the
temperature of the
ionic component reaches the
value
of the order of
$10^6\,\mathrm{K}$.
The temperatures of absorbing ions and electrons are nearly the same.
}
{b5}

\obrazekd{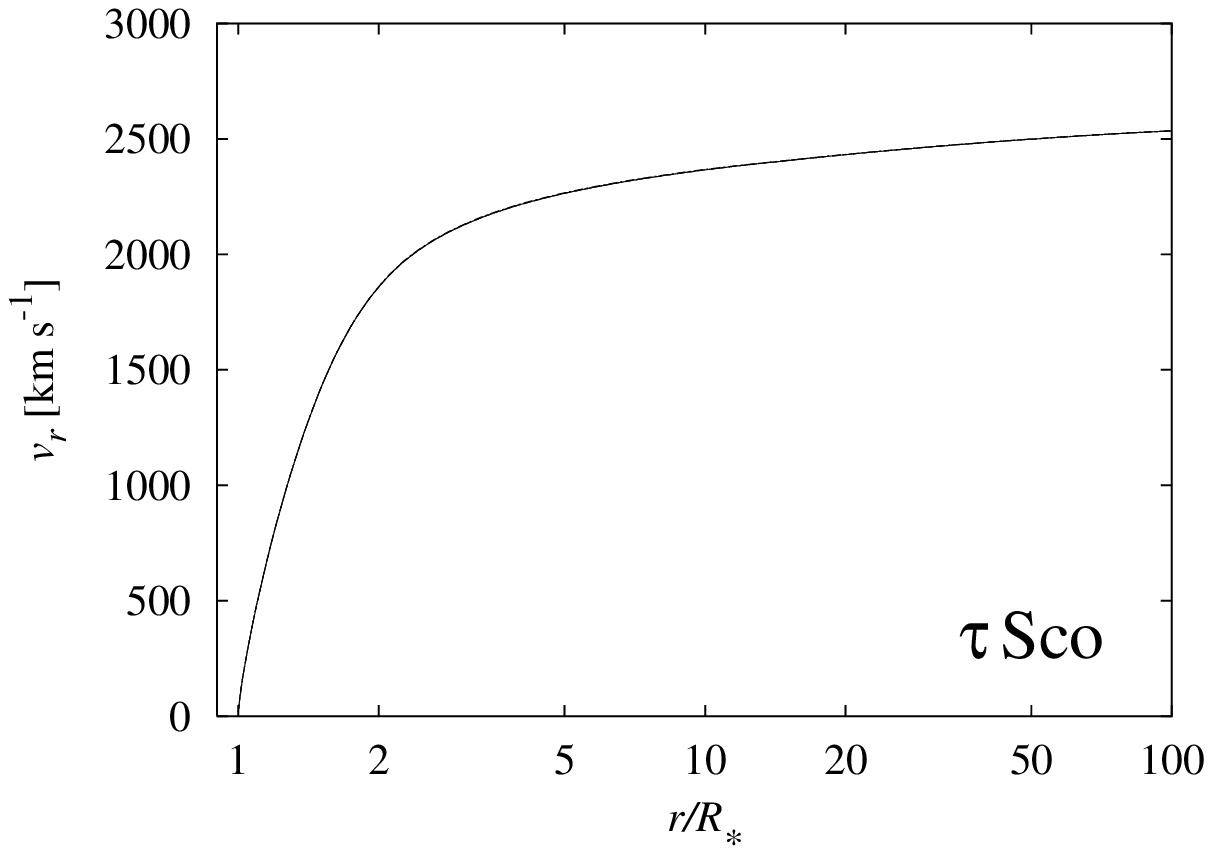}{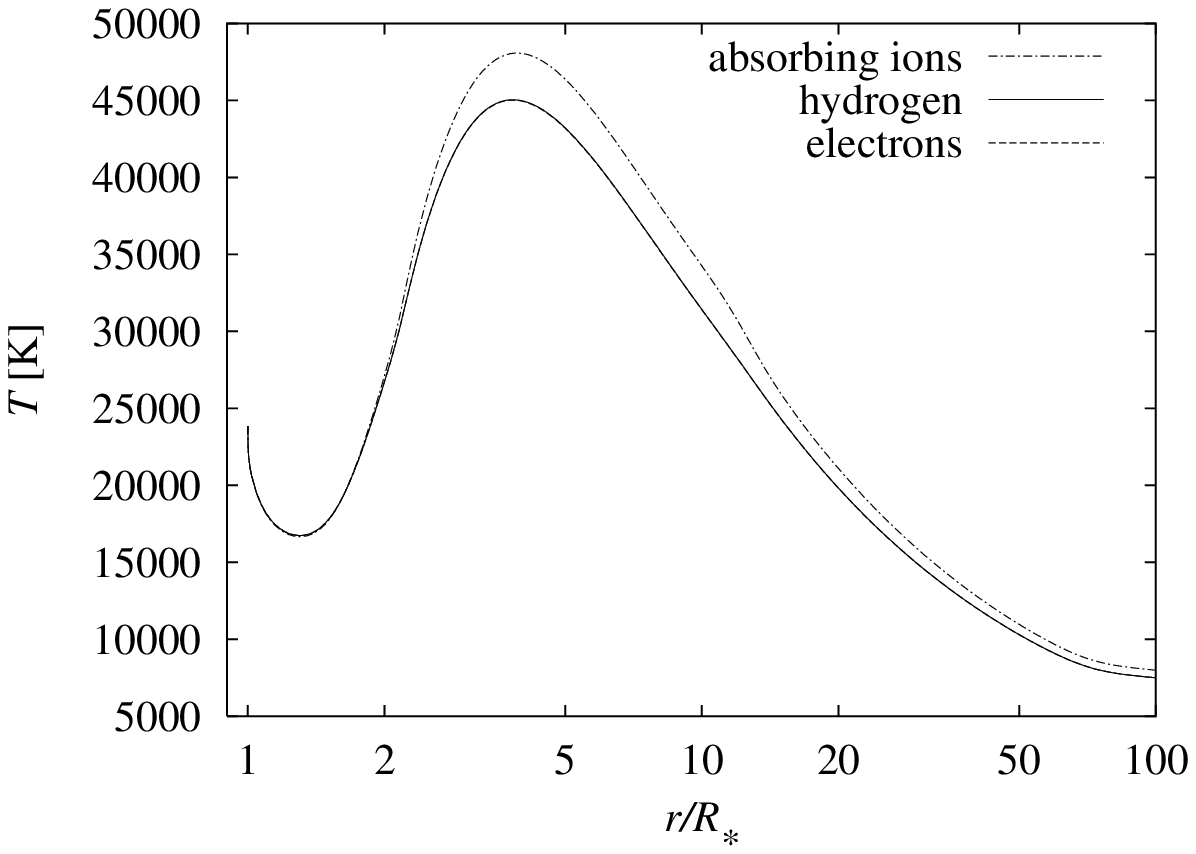}
{The same as Fig.\protect\ref{o6} for a main sequence star $\tau$~Sco.
The wind is heated both by frictional and
\GO\
heating.}
{tsco}

\obrazekd{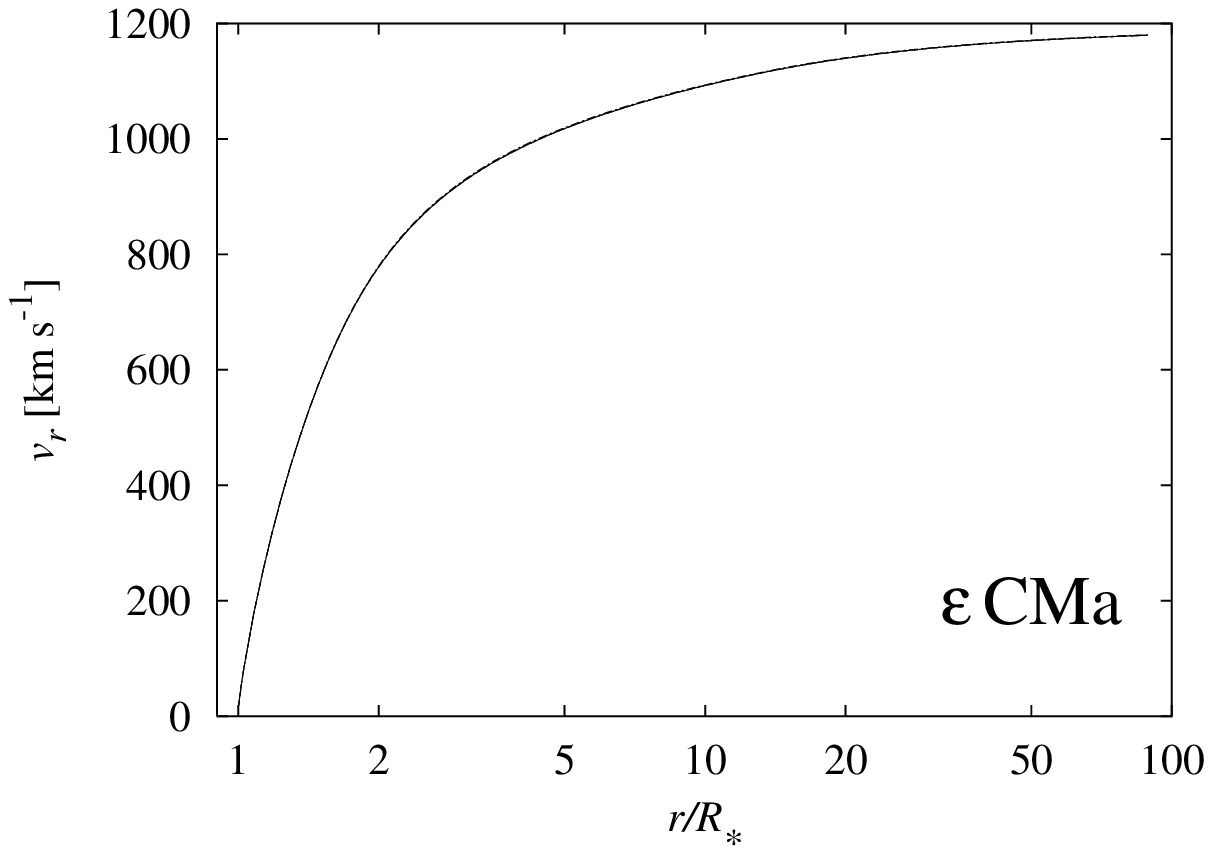}{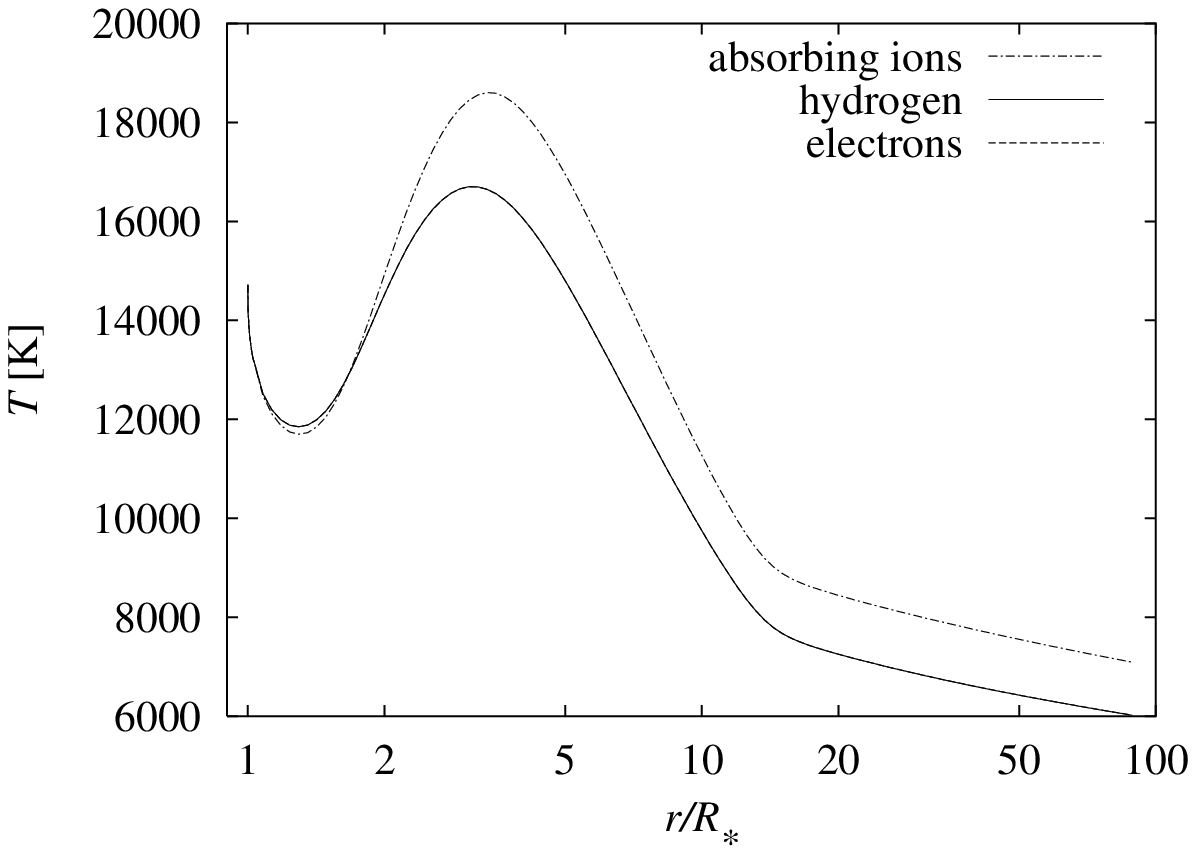}
{The same as Fig.\protect\ref{o6} for a giant star $\epsilon$~CMa.
The wind is heated both by frictional and
\GO\
heating.
Note that the wind temperature at its maximum nearly reaches the
effective temperature of the star.
}
{ecma}

\obrazekd{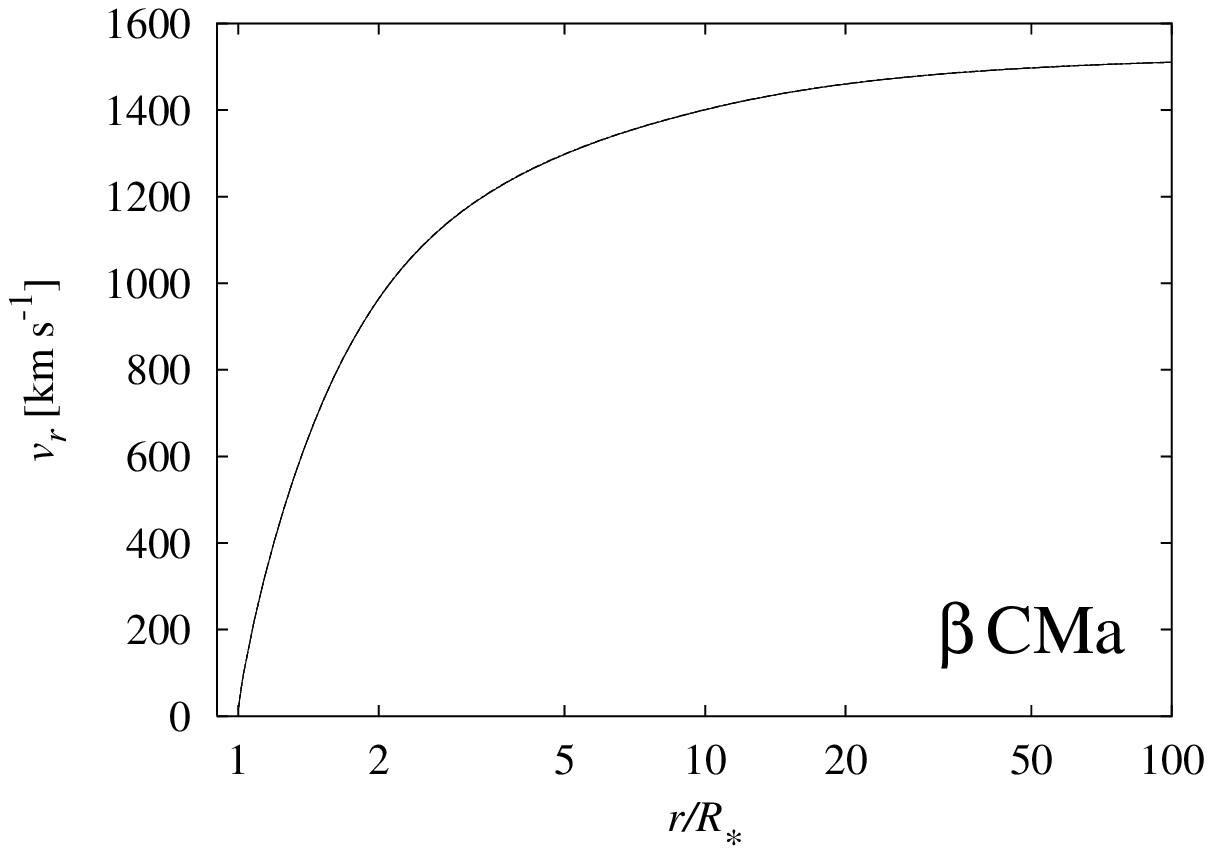}{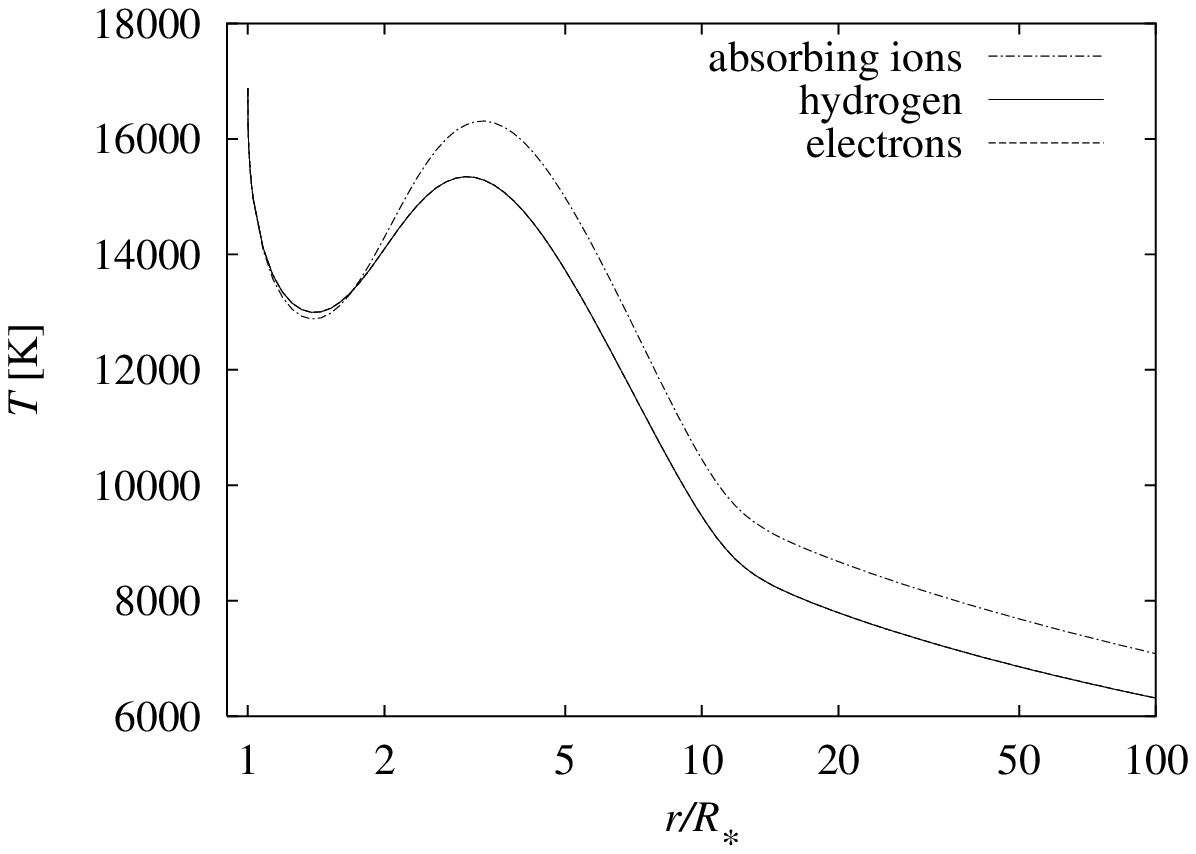}
{The same as Fig.\protect\ref{o6} for giant star 
$\beta$~CMa.
The wind is heated both by frictional and
\GO\
heating.}
{bcma}

We computed several wind models for different stellar spectral types.
Parameters of individual model stars are listed in Table~\ref{hvpar}.
Main sequence stellar parameters are taken from Harmanec (\cite{har}).
For $\tau$~Sco, $\epsilon$~CMa, and $\beta$~CMa
we used the same parameters as SP and Cassinelli et al.
(\cite{cecma}, \cite{cbcma}), respectively.
Note, that
here
we used slightly different parameters of $\tau$~Sco than KKI.
Force multipliers were adopted from Abbott (\cite{abpar}).
The parameters of absorbing ions were selected in the following way.
Because main sequence models were computed
mainly
for a demonstration of
particular effects, it was sufficient to choose an ion which is
simple enough
and which describes the basic line driving.
So,
we selected a carbon atom with $\io m=12\pr m$ as a driving ion for
them.
On the other hand,
a driving ion of individual giant models was selected more carefully
with respect to the stellar type.
For
$\epsilon$ and $\beta$ CMa we selected iron as a driving ion.
The effective temperature of $\tau$ Sco is higher, thus, we again
selected carbon as a driving ion for this star.
We stress that the selection of driving ions does not influence the
amount of radiative force.
However, it affects the thermal balance of the wind
(via the frictional and \GO\ heating).
For comparison purposes we also computed nonisothermal one-component
models (see KKI) of these stars' winds with the same stellar and wind
parameters (however, without
GO
heating).

\newlength{\staradel}
\setlength{\staradel}{\tabcolsep}
\setlength{\tabcolsep}{3.5pt}

\begin{table}[hbt]
\centering
\caption{{\em Adopted parameters of model stars.}
${\eu M}$ is the stellar mass in units of a solar mass,
$R_{*}$ is the stellar radius in units of solar radius,
$\Teff$ is the star's effective temperature,
$k$, $\alpha$, and $\delta$ are radiative force multipliers,
and $z_*$ is the metallicity.}
\label{hvpar}
\begin{tabular}{ccccccccc}
\hline
Stellar & \multicolumn{4}{c}{Stellar parameters} &
\multicolumn{3}{c}{Wind parameters} \\
type& ${\eu M}$ & $R_{*}$ & $\Teff$ & $z_*$ &
$k$ & $\alpha$ & $\delta$ & $\frac{\io m}{\pr m}$\\
 (star) & $[{\eu M}_{\odot}]$ & $[R_{\odot}]$ & $[\mathrm{K}] $ & & & & & \\
\hline
O6 & 31.65 & 9.85 & 41\,700 & 1.0 & 0.174 & 0.606 & 0.120 & 12.0 \\
O8 & 21.66 & 7.51 & 35\,600 & 1.0 & 0.166 & 0.607 & 0.120 & 12.0  \\
B0 & 14.57 & 5.80 & 29\,900 & 1.0 & 0.156 & 0.609 & 0.120 & 12.0 \\
B2 &  8.62 & 4.28 & 23\,100 & 1.0 & 0.377 & 0.537 & 0.091 & 12.0 \\
B3 &  6.07 & 3.56 & 19\,100 & 1.0 & 0.477 & 0.506 & 0.089 & 12.0 \\
B4 &  5.12 & 3.26 & 17\,200 & 1.0 & 0.365 & 0.509 & 0.105 & 12.0 \\
B5 &  4.36 & 3.01 & 15\,500 & 1.0 & 0.235 & 0.511 & 0.12 & 12.0\\
$\tau$ Sco &  19.60 & 5.50 & 33\,000 & 0.3 & 0.113 & 0.604 & 0.095& 12.0 \\
$\epsilon$ CMa & 15.2 & 16.2 & 21\,000 & 0.18 & 0.135 & 0.561 & 0.092& 55.8\\
$\beta$ CMa & 15.5 & 11.6 & 23\,250 & 0.39 & 0.125 & 0.564 & 0.099& 55.8 \\
\hline
\end{tabular}
\end{table}

\setlength{\tabcolsep}{\staradel}

\subsection{Winds with \GO\ heating}

As was shown, e.g., by SP, high density winds are well coupled.
For such winds the effect of frictional heating is negligible.
Similarly, due to a high wind density the heat exchange between
individual components is capable to maintain the same temperature for
all components.
However, for spectral types cooler than O6 a subtle effect of GO
heating/cooling influences the temperature structure.
This behaviour is displayed in Figs.\ref{o6} and \ref{o8} for wind
models of O6 and O8 stars. 
There is a large velocity gradient near the star, the variable $\sigma$
(Eq. \ref{sigma}) is positive and thus, the wind is very slightly cooled
by the GO cooling compared to the model without GO heating/cooling
effects.
On the other hand, in the outer parts of the wind the velocity gradient
is lower, the variable $\sigma$ is negative and thus GO heating
dominates (cf. Eqs. \ref{dopa}, \ref{defg}).
At the outermost parts of the wind the temperatures of models with and
without GO heating are again nearly the same mainly due to the
lowering of a stellar angular diameter.

As was discussed by GO, for stars with a lower density wind GO heating
and GO cooling are much more evident.
This can be seen in Figs.\ref{b0}, \ref{b2} for the case of B0 and B2
stars. 
As was shown by KKI, for such stars the frictional heating is negligible
and thus, changes in temperature stratification are caused only by GO
heating/cooling.
Due to the changed temperature stratification another effect becomes
important.
\zm{
Because the radiative force
in the CAK parametrization (Eq.\ref{zarzrych}) depends explicitly on the
thermal velocity,
higher wind temperature
causes its lowering.}
Lower radiative force in the outer parts of the wind (above the
critical point) leads to lowering the outflow velocity
(cf. Vink et al. \cite{bista}).
\zm{However, the description of the dependence of the radiative force on
the temperature via the
Eq.(\ref{zarzrych})
is only approximate.
Further calculations needed for better quantitative understanding the
dependence of the radiative force on the temperature are currently
under way and will be reported in future paper(s).}

\subsection{Different temperatures of individual components}

For stars with lower wind
density, individual wind components have different temperatures.
This is shown in Figs.\ref{b3} and \ref{b4} for B3 and B4 stars.

Near the stellar surface the wind is relatively dense, the heat exchange
between individual components is effective and temperatures of
individual components are nearly equal.
Compared to the model without
GO
cooling the wind temperature is
slightly lower.

However, this is not the case in the outer parts of the wind.
The wind is heated by frictional and GO heating there. 
Because the wind is more tenuous, the heat exchange between individual
components is not so effective as it is
near the star and the temperatures of individual components differ.
The temperature of absorbing ions is the highest and electron and
passive component temperatures are nearly equal. 

There are three possible mechanisms which heat (or cool) individual
components of the flow selectively and thus, which could make
temperatures of particular components different.
First, radiative heating/cooling (in our case made by bound-free and
free-free transitions) deposits (or picks up) energy from the electron
thermal pool.
This causes electrons to incline to temperatures given by former
models for whole wind -- see Drew (\cite{drewmoc}) for the one-component
case and KKI when frictional heating is included.
Second, the GO heating cools absorbing ions below the point where
$\sigma=0$ (see GO), and heats them above this point.
Finally, frictional heating itself deposits thermal energy unevenly.
From the functional behaviour of the last right-hand side term of
Eq.(\ref{rovenerg}) we can infer that temperature increase is
proportional to $\rho_b/\zav{m_a+m_b}$.
Thus, frictionally heated are mainly low-density components, i.e.
electrons and absorbing ions, both via their collisions with passive
ions. 

Another effect which influences the temperature balance
is the heat exchange between components, described by the
second right-hand side term of Eq.(\ref{rovenerg}).
Clearly, the heat exchange depends mainly on the product of number
densities of components.
Thus, similarly to the
differences in velocity,
temperature differentiation takes place mainly for the low density wind.
As discussed above, in such winds
absorbing ions can be heated more than other two components.
On the other hand,
due to their large number densities, electrons and
passive ions will share nearly the same temperature.
All these effects influence models in Fig.\ref{b3} and \ref{b4}.

We determine the charge separation field directly from the Maxwell
equation~(\ref{erov}).
However, this has only marginal effect on the wind models,
because our models tends to fulfil quasi-neutrality,
$$\el{n}\approx\pr n + \io z \io n.$$

The dependence of ionic charge for selected wind models is given in
Fig.\ref{naboj}.
Clearly, because the ionization equilibrium depends mainly on
radiative temperature,
the ionic charge is nearly constant through the wind.

Many B stars exhibit UV-excess (eg. Cassinelli et al. \cite{cecma},
\cite{cbcma}, Morales et al. \cite{moralka}).
Note, that frictional heating and
\GO\
heating
of the stellar wind could be one of the
possible explanations (Babel \cite{babela}).

\paragraph{Frictional heating in one-component models:}
The frictional heating term in the energy equation of a one-component
wind can be roughly estimated without using multicomponent models.
In the ionic equation of motion (\ref{rovhyb}) all terms without
radiative force and frictional term (corresponding to encounters between
passive component and ions)
are small.
We shall neglect them in this paragraph to derive an approximate
expression for frictional heating.
The radiative acceleration term can be expresses as
\begin{equation}
\label{rovhybjed}
{g}_{s}^{\mathrm{rad}}\approx
		\frac{1}{{\rho}_s} K_{s\mathrm{p}}G(x_{s\mathrm{p}}).
\end{equation}
Here the subscript $s$ denotes absorbing ions (note, that in these
equations we allow for more than one type of absorbing ions).
The frictional term in the energy equation of one-component wind can be
approximated as
\begin{align}
\label{trenjed}
Q^{\mathrm{fric}}&\approx\sum_{\substack{\mathrm{absorbing}\\
\mathrm{ions}}}
 K_{s\mathrm{p}}G(x_{s\mathrm{p}})|{\vr}_\mathrm{p}-{\vr}_s|\nonumber\\*
&\approx \sum_{\substack{\mathrm{absorbing}\\\mathrm{ions}}}
  {\rho}_s{g}_{s}^{\mathrm{rad}}\zav{{\vr}_s-{\vr}_\mathrm{p}},
\end{align}
where for the determination of the drift velocity from
Eq.(\ref{rovhybjed}) an approximation of the Chandrasekhar function
\begin{equation}
G(x)\approx \frac{2}{3\sqrt{\pi}}x
\end{equation}
can be used.
We computed a one-component model for a B3 star where
the frictional term (\ref{trenjed}) was inserted into the energy
equation of one-component wind model (KKI, Eq.(40c)).
Although the temperature was overestimated by $10\%$ (compared to the
correct three-component model) in the region where frictional heating is
important, the difference of the terminal velocity was only marginal. 
Note, that the frictional heating approximation Eq.(\ref{trenjed}) can
be used even when there are more than one absorbing ion component.

%%%%%%%%%%%%%%%%%%%%%%%%%%%%%%%%%%%%%%%%%%%%%%%%%%%%%%%%%%%%%%%%%%%%%%%%
\subsection{Backfalling of hydrogen to the stellar surface}

In the case of the wind with the lowest densities the absorbing ions are
not able to accelerate sufficiently the passive component of the wind.
Thus, the passive component is not dragged out of the atmosphere and
falls back onto the stellar surface (see Fig.\ref{b5} for a model of a
B5 star wind).
Such reaccretion should be studied using hydrodynamical calculations
(Porter \& Skouza \cite{obalka}).
Moreover,
Babel (\cite{babelb}) showed that 
the hydrostatic solution
for passive plasma and the wind solution for absorbing ions exists.
Probably, this type of solution is 
common for low-density winds,
because Dworetsky \& Budaj (\cite{dwobune})
whilst
studying Ne abundances in peculiar HgMn stars showed that in these stars
the radiatively driven stellar wind with hydrogen mass loss rate
larger than
$10^{-14}\,{\eu M}_{\odot}/\mathrm{yr}$ is not present.

Decoupling of velocities of absorbing and passive components is
accompanied by decoupling of temperatures of these components (see
Fig.\ref{b5}).
This effect is caused by the dependence of the amount of heat
transferred between individual components on the velocity difference
(see Eq.(\ref{rovenerg})).
Note, that the absorbing component attains temperatures sufficient to
produce X-rays.
This effect can help to explain enhanced X-ray emission observed in mid-
and late-B stars (Cohen et al. \cite{cohen}) which cannot be regarded
as a consequence of standard radiation driven wind-shock mechanism. 
Another model for X-ray emission based on the shock decoupling was
given by Porter \& Drew (\cite{iontdisk}).

\subsection{Wind models of particular stars}

As was shown by KKI, the heating effect is pronounced in the wind of
$\tau$ Sco.
Thus, we decided to recompute the wind model with the inclusion of
GO
heating.
This model is shown in the Fig.\ref{tsco}.
Similar effects as in wind models of main sequence stars occur in wind
models of giants.
This can be seen in Figs. \ref{ecma} and \ref{bcma} for the wind
models of $\beta$~CMa and $\epsilon$~CMa, respectively.
For all these stars both frictional and
GO
heating are important
for the temperature structure of the outer parts of the wind.

For the star $\tau$~Sco we used lower than observed value of
metallicity.
Contrary to Kilian (\cite{kil}) who determined the value $z_*=0.6$ we
reduced the metallicity to $z_*=0.3$
to enable larger frictional heating.
This change reflects mainly uncertainties of our model, because, e.g.,
our model with $z_*=0.5$ and metallic component described by iron ions
instead carbon ions yields nearly the same velocity and temperature
stratification.

Unfortunately, existing measurements of the terminal velocity for this
star do not allow to verify our models precisely.
Abbott (\cite{abvnek}) and Lamers \& Rogerson (\cite{lamrog}) determined
$v_{\infty}=2000\,\mathrm{km}\,\mathrm{s}^{-1}$, whereas Lamers et al.
(\cite{lsl}) measured $v_{\infty}=1000\,\mathrm{km}\,\mathrm{s}^{-1}$.
However, all of them claim that their values are uncertain.
Larger values of $v_{\infty}$ are supported also by a detailed UV-fit of
Hamann (\cite{ham}).

For $\epsilon$~CMa we used metallicity $z_*=0.18$, a value estimated by
Gies \& Lambert (\cite{gilam}).
Similarly to $\tau$~Sco, available determinations of terminal velocity
have lower quality.
Abbott (\cite{abvnek}) determined
$v_{\infty}=700\,\mathrm{km}\,\mathrm{s}^{-1}$
and Lamers et al. (\cite{lsl}) measured
$v_{\infty}=600\,\mathrm{km}\,\mathrm{s}^{-1}$.
However, it is not clear whether the apparent discrepancy of theoretical
and observational terminal velocities is caused by the models or is due
to the 
inaccurate
measurements.

According to Gies \& Lambert (\cite{gilam}) we reduced the metallicity
of $\beta$~CMa to the value $z_*=0.39$.
To our knowledge, there is no measured terminal velocity for this
star available in the literature.
Note that for both $\epsilon$~CMa and $\beta$~CMa
enhanced wind temperature can help to explain observed
UV-excess (Cassinelli et al. \cite{cecma}, \cite{cbcma}).

%%%%%%%%%%%%%%%%%%%%%%%%%%%%%%%%%%%%%%%%%%%%%%%%%%%%%%%%%%%%%%%%%%%%%%%%
\section{Comparison of terminal velocities}

\begin{table*}[hbt]
\begin{center}
\caption{Stellar and wind parameters of O6 -- B5 stars selected from
LSL.
$\dot{\eu M}$ is computed mass-loss rate.
Terminal velocities measured by LSL (column $v_{\infty}$ (LSL))
are compared with theoretical values obtained by LSL using a ``cooking
formula'' of KPPA (column $v_{\infty}$ (KPPA)) and with predicted ones
computed with an assumption of a nonisothermal wind model (column
$v_{\infty}$ (predicted)).}
\label{tablsl}
\begin{tabular}{cccccccccccc}
\hline
HD & \multicolumn{4}{c}{Stellar parameters} &
\multicolumn{3}{c}{Wind parameters}& & \multicolumn{3}{c}{Terminal velocities} \\
number& ${\eu M}$ & $R_{*}$ & $\Teff$ & $z_*$ &
$k$ & $\alpha$ & $\delta$ & $\dot{\eu M}$ & $v_{\infty}$ (LSL)  & $v_{\infty}$ (KPPA)  &
$v_{\infty}$ (predicted)\\
 & $[{\eu M}_{\odot}]$ & $[R_{\odot}]$ & $[\mathrm{K}] $ & & & & & 
$[{\eu M}_{\odot}\,\mathrm{yr}^{-1}]$ &
$[\mathrm{km}\,\mathrm{s}^{-1}]$ & $[\mathrm{km}\,\mathrm{s}^{-1}]$ &
$[\mathrm{km}\,\mathrm{s}^{-1}]$ \\
\hline
$30614$ & $43.0$ & $27.6$ & $30900$ & $1.0$ & $0.158$ & $0.609$ & $0.120$ &  
$\zm{8.9\;10^{-6}}$& $ 1500\,\pm\,200 $ & $2241$ & $1450$ \\
$ 34656 $ & $ 30.0 $ & $ 9.9 $ & $ 38100 $ & $ 1.0 $ & $ 0.171 $ & $ 0.607 $ & $ 0.120 $ & 
 $\zm{1.0\;10^{-6}}$ &$ 2100\,\pm\,100 $ & $3598$ & $2590$ \\
$36861$ & $30.0$ & $12.3$ & $36000$ & $1.0$ & $0.167$ & $0.607$ & $0.120$ &  
$\zm{1.5\;10^{-6}}$ &$ 2200\,\pm\,300 $ & $3084$ & $2230$ \\
$41117$ & $25.0$ & $43.4$ & $18500$ & $1.0$ & $0.410$ & $0.507$ & $0.098$ &  
$\zm{2.8\;10^{-6}}$ &$ 500\,\pm\,50 $ & $1058$ & $740$ \\
$43384$ & $19.0$ & $39.8$ & $16300$ & $1.0$ & $0.311$ & $0.510$ & $0.112$ &  
$\zm{4.0\;10^{-7}}$ &$ 500\,\pm\,100 $ & $1015$ & $690$ \\
$47240$ & $17.0$ & $23.4$ & $20800$ & $1.0$ & $0.451$ & $0.514$ & $0.091$ &  
$\zm{9.7\;10^{-7}}$ &$ 1000\,\pm\,100 $ & $1267$ & $930$ \\
$51309$ & $11.0$ & $16.3$ & $16700$ & $1.0$ & $0.329$ & $0.509$ & $0.109$ &  
$\zm{1.7\;10^{-8}}$ &$ 700\,\pm\,100 $ & $1309$ & $910$ \\
$52382$ & $17.0$ & $20.4$ & $20800$ & $1.0$ & $0.451$ & $0.514$ & $0.091$ &  
$\zm{4.9\;10^{-7}}$ &$ 1200\,\pm\,100 $ & $1381$ & $1030$ \\
$69464$ & $49.0$ & $20.1$ & $37200$ & $1.0$ & $0.169$ & $0.607$ & $0.120$ &  
$\zm{9.9\;10^{-6}}$ &$ 2100\,\pm\,200 $ & $2721$ & $1840$ \\
$74194$ & $28.0$ & $14.5$ & $33000$ & $1.0$ & $0.161$ & $0.608$ & $0.120$ &  
$\zm{1.4\;10^{-6}}$ &$ 2000\,\pm\,300 $ & $2829$ & $1970$ \\
$79186$ & $18.0$ & $62.4$ & $13600$ & $1.0$ & $0.284$ & $0.519$ & $0.100$ &  
$\zm{6.6\;10^{-7}}$ &$ 450\,\pm\,50 $ & $772$ & $530$ \\
$91572$ & $38.0$ & $9.6$ & $42200$ & $1.0$ & $0.175$ & $0.606$ & $0.114$ & 
$\zm{1.6\;10^{-6}}$ &$ 2400\,\pm\,100 $ & $3989$ & $2940$ \\
$91969$ & $25.0$ & $22.9$ & $26000$ & $1.0$ & $0.284$ & $0.568$ & $0.108$ &  
$\zm{3.0\;10^{-6}}$ &$ 1500\,\pm\,100 $ & $1816$ & $1240$ \\
$92964$ & $29.0$ & $68.4$ & $17400$ & $1.0$ & $0.361$ & $0.509$ & $0.105$ &  
$\zm{1.1\;10^{-5}}$ &$ 550\,\pm\,50 $ & $807$ & $530$ \\
$93130$ & $43.0$ & $13.8$ & $40200$ & $1.0$ & $0.174$ & $0.606$ & $0.119$ &  
$\zm{4.5\;10^{-6}}$ &$ 2500 \,\pm\, 300 $ & $3337$ & $2370$ \\
$96248$ & $25.0$ & $38.9$ & $20800$ & $1.0$ & $0.451$ & $0.514$ & $0.091$ &  
$\zm{7.3\;10^{-6}}$ &$ 650\,\pm\,50 $ & $1116$ & $770$ \\
$96917$ & $46.0$ & $25.2$ & $33000$ & $1.0$ & $0.161$ & $0.608$ & $0.120$ &  
$\zm{9.5\;10^{-6}}$ &$ 1800\,\pm\,200 $ & $2430$ & $1600$ \\
$101190$ & $48.0$ & $13.9$ & $42200$ & $1.0$ & $0.175$ & $0.606$ & $0.114$ & 
$\zm{6.3\;10^{-6}}$ &$ 2800\,\pm\,200 $ & $3452$ & $2450$ \\
$101436$ & $42.0$ & $12.4$ & $41200$ & $1.0$ & $0.174$ & $0.606$ & $0.117$ &  
$\zm{3.5\;10^{-6}}$ &$ 2700\,\pm\,200 $ & $3487$ & $2570$ \\
$106343$ & $24.0$ & $40.7$ & $19700$ & $1.0$ & $0.464$ & $0.506$ & $0.091$ &  
$\zm{5.3\;10^{-6}}$ &$ 800\,\pm\,100 $ & $1074$ & $730$ \\
$109867$ & $26.0$ & $38.9$ & $20800$ & $1.0$ & $0.451$ & $0.514$ & $0.091$ &  
$\zm{6.7\;10^{-6}}$ &$ 1200\,\pm\,200 $ & $1144$ & $800$ \\
$112244$ & $46.0$ & $25.2$ & $33000$ & $1.0$ & $0.161$ & $0.608$ & $0.120$ &  
$\zm{9.5\;10^{-6}}$ &$ 1600\,\pm\,100 $ & $2430$ & $1580$ \\
$116084$ & $15.0$ & $24.8$ & $17400$ & $1.0$ & $0.361$ & $0.509$ & $0.105$ &  
$\zm{1.4\;10^{-7}}$ &$ 500\,\pm\,100 $ & $1201$ & $830$ \\
$148379$ & $24.0$ & $40.7$ & $19700$ & $1.0$ & $0.464$ & $0.506$ & $0.091$ &  
$\zm{5.3\;10^{-6}}$ &$ 500\,\pm\,100 $ & $1074$ & $730$ \\
$151515$ & $41.0$ & $14.9$ & $38100$ & $1.0$ & $0.171$ & $0.607$ & $0.120$ &  
 $\zm{4.1\;10^{-6}}$ &$ 2400\,\pm\,100 $ & $3196$ & $2230$ \\
$151804$ & $70.0$ & $34.0$ & $34000$ & $1.0$ & $0.163$ & $0.608$ & $0.120$ &  
$\zm{3.6\;10^{-5}}$ &$ 1500\,\pm\,200 $ & $2226$ & $1370$ \\
$152405$ & $25.0$ & $15.3$ & $30500$ & $1.0$ & $0.157$ & $0.609$ & $0.120$ &  
$\zm{1.0\;10^{-6}}$ &$ 1800\,\pm\,200 $ & $2616$ & $1800$ \\
$152424$ & $52.0$ & $33.4$ & $30500$ & $1.0$ & $0.157$ & $0.609$ & $0.120$ &  
$\zm{1.5\;10^{-5}}$ &$ 1500\,\pm\,100 $ & $2056$ & $1350$ \\
$154090$ & $26.0$ & $38.9$ & $20800$ & $1.0$ & $0.451$ & $0.514$ & $0.091$ &  
$\zm{6.7\;10^{-6}}$ &$ 950\,\pm\,50 $ & $1144$ & $800$ \\
$157246$ & $17.0$ & $23.4$ & $20800$ & $1.0$ & $0.451$ & $0.514$ & $0.091$ &  
$\zm{9.7\;10^{-7}}$ &$ 900\,\pm\,200 $ & $1267$ & $930$ \\
$162978$ & $40.0$ & $16.0$ & $37100$ & $1.0$ & $0.169$ & $0.607$ & $0.120$ &  
$\zm{4.4\;10^{-6}}$ &$ 2200\,\pm\,200 $ & $2939$ & $2090$ \\
$163758$ & $50.0$ & $20.1$ & $37200$ & $1.0$ & $0.169$ & $0.607$ & $0.120$ &  
$\zm{9.5\;10^{-6}}$ &$ 2300\,\pm\,200 $ & $2765$ & $1890$ \\
$166596$ & $9.7$ & $9.8$ & $18700$ & $1.0$ & $0.419$ & $0.507$ & $0.097$ & 
$\zm{9.6\;10^{-9}}$ &$ 700\,\pm\,50 $ & $1551$ & $1170$ \\
$175754$ & $34.0$ & $14.2$ & $36000$ & $1.0$ & $0.167$ & $0.607$ & $0.120$ &  
$\zm{2.4\;10^{-6}}$ &$ 2000\,\pm\,100 $ & $2998$ & $2130$ \\
$186980$ & $35.0$ & $13.9$ & $37100$ & $1.0$ & $0.169$ & $0.607$ & $0.120$ &  
$\zm{2.8\;10^{-6}}$ &$ 2100\,\pm\,100 $ & $3028$ & $2180$ \\
$188209$ & $43.0$ & $27.6$ & $30900$ & $1.0$ & $0.158$ & $0.609$ & $0.120$ &  
$\zm{8.9\;10^{-6}}$ &$ 1600\,\pm\,100 $ & $2241$ & $1450$ \\
$190603$ & $24.0$ & $40.7$ & $19700$ & $1.0$ & $0.464$ & $0.506$ & $0.091$ &  
$\zm{5.3\;10^{-6}}$ &$ 500\,\pm\,50 $ & $1074$ & $730$ \\
$190864$ & $42.0$ & $14.0$ & $39200$ & $1.0$ & $0.173$ & $0.606$ & $0.120$ &  
$\zm{3.8\;10^{-6}}$ &$ 2300\,\pm\,200 $ & $3338$ & $2340$ \\
$198478$ & $17.0$ & $36.3$ & $16300$ & $1.0$ & $0.311$ & $0.510$ & $0.112$ &  
$\zm{3.0\;10^{-7}}$ &$ 550\,\pm\,100 $ & $1009$ & $690$ \\
$204172$ & $23.0$ & $20.0$ & $26000$ & $1.0$ & $0.284$ & $0.568$ & $0.108$ &  
$\zm{1.8\;10^{-6}}$ &$ 1600\,\pm\,200 $ & $1911$ & $1320$ \\
$206165$ & $19.0$ & $35.8$ & $18500$ & $1.0$ & $0.410$ & $0.507$ & $0.098$ &  
$\zm{1.6\;10^{-6}}$ &$ 600\,\pm\,50 $ & $1011$ & $720$ \\
$210809$ & $38.0$ & $21.4$ & $32000$ & $1.0$ & $0.160$ & $0.608$ & $0.120$ &  
$\zm{4.2\;10^{-6}}$ &$ 2000\,\pm\,200 $ & $2588$ & $1700$ \\
$210839$ & $51.0$ & $19.6$ & $38200$ & $1.0$ & $0.171$ & $0.607$ & $0.120$ &  
$\zm{1.1\;10^{-5}}$ &$ 2200 \,\pm\, 200 $ & $2882$ & $1930$ \\
$213087$ & $21.0$ & $23.4$ & $23400$ & $1.0$ & $0.368$ & $0.541$ & $0.100$ &  
$\zm{1.9\;10^{-6}}$ &$ 1400\,\pm\,200 $ & $1533$ & $1080$ \\
$218915$ & $43.0$ & $27.6$ & $30900$ & $1.0$ & $0.158$ & $0.609$ & $0.120$ &  
$\zm{8.9\;10^{-6}}$ &$ 1800\,\pm\,100 $ & $2241$ & $1450$ \\
\hline
\end{tabular}
\end{center}
\end{table*}

\obrazek{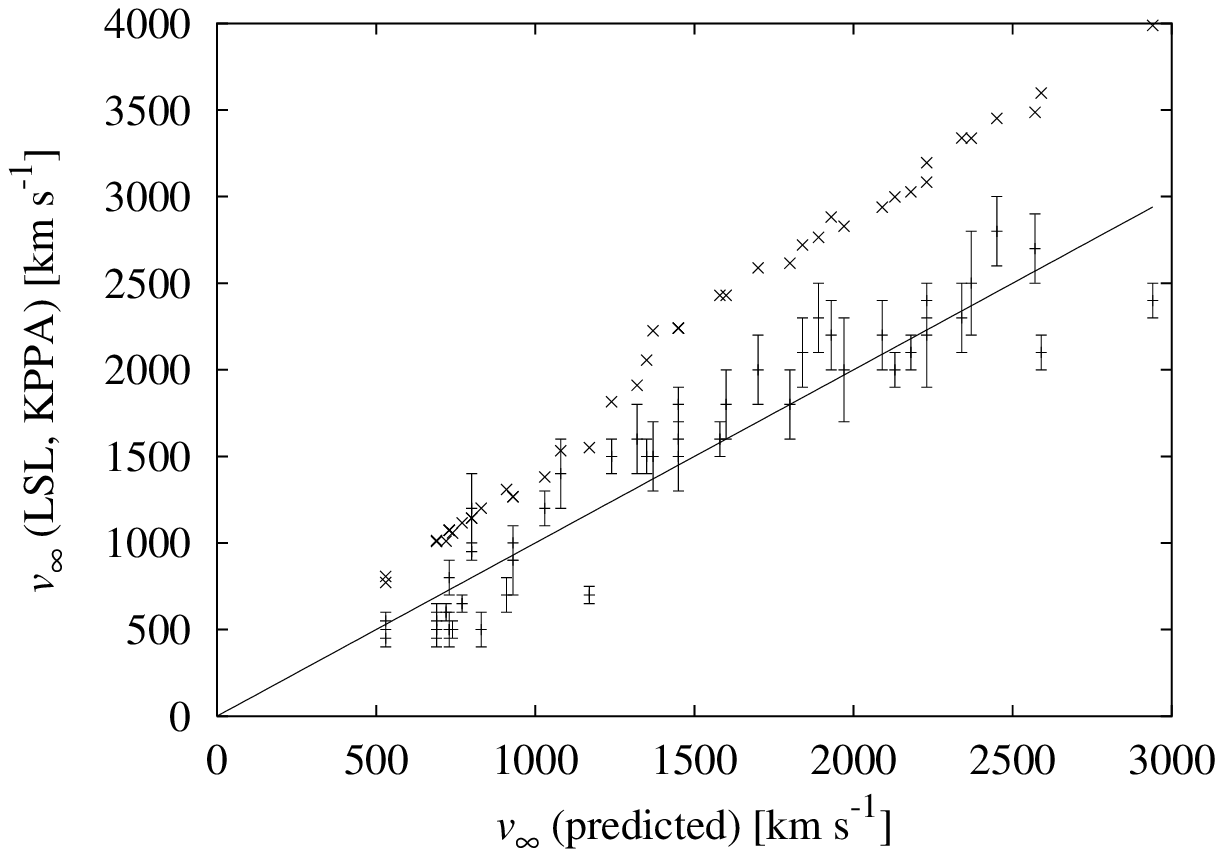}{Comparison of predicted and observational (taken
from LSL) values of terminal velocities.
Vertical lines denote uncertainty of observed values. Straight line is
one-to-relation.
For comparison, we plotted theoretical terminal velocities computed by
LSL using ``cooking formula'' (crosses).}{porlsl}

In addition, we decided to compare our predicted terminal velocities
with that measured by Lamers et al. (\cite{lsl}, hereafter LSL).
They found discrepancy between theoretical values obtained from a
``cooking formula'' of Kudritzki et al. (\cite{varic}, hereafter KPPA) 
and their experimental values.
We computed wind models of O6 -- B5 stars for which LSL measured the
terminal velocity.
Parameters of each wind model are given in Tab. \ref{tablsl}.
Stellar parameters are taken from LSL, wind parameters are adopted from
Abbott~(\cite{abpar}).
For many stars we found quite a good agreement between observed and
predicted terminal velocities (see Fig.\ref{porlsl} for comparison of
predicted and observed values).
Although some values of predicted terminal velocities miss the measured
value significantly (e.g. for the star HD\,166596), it is evident that
the overall agreement between our predicted terminal velocities and the
observed ones is {\em much better} than that of the ``cooking formula''
of KPPA and the systematic difference,
\zm{which was previously attributed to an overestimation of $\alpha$
(by LSL),}
has been removed.

\zm{However, there are still differences between observed and predicted
values of $v_\infty$.}
There are three basic reasons for
this discrepancy.
First, rotation
lowers the terminal velocity (cf. Friend \& Abbott \cite{fa}).
However, Petrenz \& Puls (\cite{ppjet}) using 2D models showed that the
influence of the rotation on terminal velocity in many cases is only
marginal.
Second, our wind models (especially the radiative force) are
constrained.
Although
we included physical processes that have not been
included yet (frictional heating,
\GO\
heating, multicomponent
nature of the wind), there are still limits.
Our treatment of ionization is only approximate, the equilibrium is
not determined consistently with radiation field.
In addition, our models are not fully consistent with respect to the
radiative force, a proper NLTE treatment of the radiative transfer
problem would be very useful.
This two reasons causes that many of the terminal velocities are not
within quoted uncertainties. 
However, we plan to improve our models in near future.

Another source of differences may come from uncertainties of stellar
parameters derived from observations.
Note that, e.g., stars HD\,106343, HD\,148379, and HD\,190603 have
fairly the same parameters, however different observed terminal
velocities.

The ``cooking formula'' of KPPA should be consistent with detailed
calculations of Pauldrach et al. (\cite{ppk})
{\zm with an accuracy about 5\%.}
However, our predicted terminal velocities correspond to those
computed by Pauldrach et al. (\cite{ppk}), too.
Thus, there is not clear source of discrepancy between terminal
velocities observed by LSL and predicted using formula of KPPA.
We stress that the effect of frictional or \GO\ heating on the
terminal velocity are negligible
for the models described in this section.

\section{Conclusions}

We computed non-isothermal three-component models of OB star winds with
allowing for
different temperatures of each component and with inclusion of the
\GO\ (GO)
heating/cooling.
We showed that temperature differentiation takes place in the winds of
B stars starting from spectral type B3.
The temperature of absorbing ions is of the order $10^3\,\mathrm{K}$
higher than temperature of other components whereas the temperatures of
passive plasma and electrons is nearly equal.
The main sources which trigger the temperature differentiation are
GO,
frictional, and radiative heating.

Another important effect studied in this paper is the GO heating
and cooling, which is important mainly for the low density winds.
We showed that this effect is a direct consequence of the dependence of
the radiative force on the wind velocity.
We derived the GO heating formula directly from the Boltzmann equation.
More subtle GO cooling operates near the star at the wind base whereas
the GO heating affects the flow mainly in outer parts of the wind.
These effects become important starting from stellar type O6.
Frictional and GO heating provides a possibility for an alternative
explanation of UV-excess observed in some B stars.

At the lowest densities either the passive component falls back onto the
star or purely metallic wind exists.
If the reaccretion takes place then ionic components is frictionally
heated to the temperatures of orders millions~K creating corona-like
region.
This effect can explain enhanced X-ray activity in many of B stars.

\zm{Finally, we}
compared our computed terminal velocities with that derived from
observation.
There is quite good agreement between them.
The systematic difference between observed and predicted (by a
``cooking formula'' of KPPA) terminal velocities found by LSL was
removed.
However, we found no effects of frictional or GO heating in our sample.

\begin{acknowledgements}
The authors would like to thank Dr. John Porter for pointing their
attention to the importance of the effect of \GO\ (Doppler) heating,
Dr. Kenneth Gayley and Prof. Michal Lenc for their comments to the
manuscript of this paper.
This research has made use of NASA's Astrophysics Data System Abstract
Service (\cite{ADS1}, \cite{ADS2}, \cite{ADS3}, \cite{ADS4}). 
This work was supported by a grant GA \v{C}R 205/01/0656 and by
projects K2043105 and Z1003909.
\end{acknowledgements}

\appendix

\section{Numerical calculation of $G(\sigma,\mu_*)$}
\label{vypg}

The function $G(\sigma,\mu_*)$ is computed using numerical quadrature.
Firstly, integral over $x$ can be efficiently computed using a
Hermite quadrature formula (cf. Ralston \cite{ral}).
Quadrature weights and knots
were computed using a subroutine {\tt  IQPACK} which is an
implementation of method described by Kautsky \& Elhay (\cite{iq}).
Satisfactory approximation can be obtained using 20 quadrature points.

For an angle integration we used Legendre quadrature formula with 5
quadrature points.
Again, quadrature weights and knots were computed using subroutine
{\tt  IQPACK} (Kautsky \& Elhay \cite{iq}).

Finally, the integration over $y$ was performed using Simpson quadrature
rule.
The quadrature integral $(0,\infty)$ was approximated by
$(10^{-3},10^{5})$ and the Simpson integration was divided into
subintervals of power 10, with 10 quadrature points in each of them. 

Numerical tests showed that temperature computed with described
approximation of the
GO
heating has an error less than $1\%$.

\end{document}